\documentclass[aps,prl,twocolumn,superscriptaddress]{revtex4}
\usepackage{bm,bbold}
\usepackage[utf8]{inputenc}
\usepackage{xcolor}
\usepackage{amsmath}
\usepackage{amssymb}
\usepackage{graphicx}
\usepackage{braket}
\usepackage{float}
\usepackage{psfrag}

\newcommand{\bk}{\mathbf{k}}

\newcommand{\bp}{\mathbf{p}}
\newcommand{\bq}{\mathbf{q}}
\newcommand{\bQ}{\mathbf{Q}}
\def\dz2{d$_{\text{z}^2}$}

\def\dx2y2{d$_{\text{x}^2\text{y}^2}$}
\def\G0W0{G$_0$W$_0$}
\def\scGW0{scGW$_0$}

\begin{document}

\author{A. Steinhoff}
%\affiliation[Institut f\"ur Theoretische Physik]
\affiliation{Institut f\"ur Theoretische Physik, Universit\"at Bremen, P.O. Box 330 440, 28334 Bremen, Germany}
%\email{asteinhoff@itp.uni-bremen.de}

\author{M. Florian}
%\affiliation[Institut f\"ur Theoretische Physik]
\affiliation{Institut f\"ur Theoretische Physik, Universit\"at Bremen, P.O. Box 330 440, 28334 Bremen, Germany}

\author{M. R\"osner}
%\affiliation[ITP]
\affiliation{Institut f\"ur Theoretische Physik, Universit\"at Bremen, P.O. Box 330 440, 28334 Bremen, Germany}
%\alsoaffiliation[Bremen Center for Computational Materials Science]
\affiliation{Bremen Center for Computational Materials Science, Universit\"at Bremen, 28334 Bremen, Germany}

\author{G. Sch\"onhoff}
%\affiliation[ITP]
\affiliation{Institut f\"ur Theoretische Physik, Universit\"at Bremen, P.O. Box 330 440, 28334 Bremen, Germany}
%\alsoaffiliation[Bremen Center for Computational Materials Science]
\affiliation{Bremen Center for Computational Materials Science, Universit\"at Bremen, 28334 Bremen, Germany}

\author{T.O. Wehling}
%\affiliation[ITP]
\affiliation{Institut f\"ur Theoretische Physik, Universit\"at Bremen, P.O. Box 330 440, 28334 Bremen, Germany}
%\alsoaffiliation[Bremen Center for Computational Materials Science]
\affiliation{Bremen Center for Computational Materials Science, Universit\"at Bremen, 28334 Bremen, Germany}

\author{F. Jahnke}
%\affiliation[Institut f\"ur Theoretische Physik]
\affiliation{Institut f\"ur Theoretische Physik, Universit\"at Bremen, P.O. Box 330 440, 28334 Bremen, Germany}

%%%%%%%%%%%%%%%%%%%%%%%%%%%%%%%%%%%%%%%%%%%%%%%%%%%%%%%%%%%%%%%%%%%%%
%% The document title should be given as usual. Some journals require
%% a running title from the author: this should be supplied as an
%% optional argument to \title.
%%%%%%%%%%%%%%%%%%%%%%%%%%%%%%%%%%%%%%%%%%%%%%%%%%%%%%%%%%%%%%%%%%%%%
% \title{Excitons versus plasma in monolayer TMDC semiconductors: Ionization equilibrium and Mott transition}
% \title{Exciton ionization equilibrium and Mott transition in monolayer transition metal dichalcogenide semiconductors}
\title{Excitons versus electron-hole plasma in monolayer transition metal dichalcogenide semiconductors}

\begin{abstract}

\textbf{When electron-hole pairs are excited in a semiconductor, it is a priori not clear if they form a fermionic plasma of unbound particles or a bosonic exciton gas. 
Usually, the exciton phase is associated with low temperatures.
In atomically thin transition metal dichalcogenide semiconductors, excitons are particularly important even 
at room temperature due to strong Coulomb interaction and a large exciton density of states. 
Using state-of-the-art many-body theory including dynamical screening, we show that
the exciton-to-plasma ratio can be efficiently tuned by dielectric substrate screening as well as charge carrier doping. 
Moreover, we predict a Mott transition from the exciton-dominated regime to a fully ionized electron-hole plasma at excitation densities 
between $3\times10^{12}$~cm$^{-2}$ and $1\times10^{13}$~cm$^{-2}$ depending on temperature, carrier doping and dielectric environment. 
We propose the observation of these effects by studying
excitonic satellites in photoemission spectroscopy and scanning tunneling microscopy.
}
% \\
% \\ \textbf{Keywords: excitons, plasmons, dichalcogenides, 2D materials, Coulomb interaction, many-particle effects}

\end{abstract}

\maketitle

%%%%%%%%%%%%%%%%%%%%%%%%%%%%%%%%%%%%%%%%%%%%%%%%%%%%%%%%%%%%%%%%%%%%%
%% INTRODUCTORY PART
%%%%%%%%%%%%%%%%%%%%%%%%%%%%%%%%%%%%%%%%%%%%%%%%%%%%%%%%%%%%%%%%%%%%%

% \section{Introduction.} 

Excitons play a prominent role in the optical properties of atomically thin transition metal dichalcogenide (TMDC) semiconductors due to electron-hole Coulomb interaction
being strongly enhanced by carrier confinement and reduced dielectric screening.
% This is why oftentimes interpretation of experimental results as well as theoretical prediction takes place
This suggests an interpretation of experimental results as well as theoretical prediction
in terms of excitons rather than unbound
electrons and holes. \cite{zhang_absorption_2014,berghauser_analytical_2014,sun_observation_2014,schmidt_ultrafast_2016,selig_excitonic_2016} On the other hand, it is well-known that at a certain
excitation density of electron-hole pairs the Mott transition is observed. \cite{shah_investigation_1977,manzke_mott_2012,chernikov_electrical_2015} 
Here a phase where excitons and unbound carriers can coexist evolves into a fully ionized electron-hole plasma, which shows the significance of plasma effects to the 
physics of excited semiconductors.
% \rem{It has also been shown in a recent experiment that excitons and plasma coexist after non-resonant exitation of monolayer WSe$_2$ \cite{steinleitner_direct_2017}}
% involves plasma effects as well and 

Since excitons are more or less neutral compound particles, many-particle renormalization and screening effects in an exciton gas are very different from those in a plasma of 
unbound electrons and holes, which we refer to in the following as quasi-free carriers. For this reason it is highly desirable to quantify the relative importance of excitonic 
and plasma effects over a wide range of electron-hole excitation densities and for various material and externally controllable parameters. %in order
% This makes for understanding the fundamental properties of excited TMDC semiconductors and utilizing them for device application, e.g. in optoelectronics. 
It has already been suggested to tune 
exciton binding energies by electrical doping \cite{chernikov_electrical_2015} and some effort has been devoted to study the influence of dielectric screening
on excitons in TMDC semiconductors. \cite{lin_dielectric_2014,ugeda_giant_2014,latini_excitons_2015,trolle_model_2017}

% -interesting to know how to describe system: either by excitonic [Ref: Berlin, Marburg] or plasma approaches [Ref] and when mott transition takes place

% -interpretation of optical spectra relies on exciton/plasma
% 
% -X screen less, as we show in this paper

In the past, a very powerful scheme has been developed to theoretically describe the so-called ionization equilibrium between the formation of bound particles and 
their dissociation into quasi-free particles \cite{kremp_equation_1984,kremp_ladder_1984,zimmermann_mass_1985,kremp_quantum_1993, kremp_quantum_2005, semkat_ionization_2009}, 
with applications to atomic plasmas and highly excited semiconductors.  
% Although the ionization equilibrium is very sensitive to a correct treatment of many-body interaction effects, it can be understood to a large degree 
% from a simple chemical picture: Here, excitons are interpreted as a new particle species with a chemical potential of their own that equilibrate with electrons and holes as in a chemical reaction.
In general, a dominance of free-carrier plasma over excitons is expected at very low excitation density due to entropy ionization of excitons \cite{mock_entropy_1978} and above the so-called Mott density due to 
pressure ionization of excitons. The fraction of excitons reaches a maximum at elevated densities still below the Mott transition and can assume values of practically $100\%$ in TMDCs, though being sensitive to
various externally controllable parameters.
The scheme relies on the assumption of a quasi-equilibrium between plasma and excitons being established before electron-hole recombination sets in.
% too many excited carriers have recombined either radiatively or non-radiatively. 
Ultrafast equilibration is facilitated by efficient carrier-carrier \cite{steinhoff_nonequilibrium_2016} and carrier-phonon interaction \cite{selig_excitonic_2016} 
as well as exciton formation \cite{ceballos_exciton_2016,steinleitner_direct_2017} after optical excitation, see Ref.~\citenum{koch_semiconductor_2006} for a review.
% This might be prepared by quasi-resonant 
% optical excitation generating a coherent excitonic polarization that decays via efficient carrier-phonon scattering into incoherent exciton populations. 
% Via subsequent scattering processes, the exciton population relaxes towards an equilibrium distribution according to the actual temperature. Alternatively, non-resonant excitation above the single-particle 
% band gap generates an incoherent electron-hole plasma that equilibrates on a sub-$100$-fs timescale due to carrier-carrier \cite{steinhoff_nonequilibrium_2016} and carrier-phonon interaction [Ref] and consequently 
% forms an exciton population. 
% After formation, bright excitons with vanishing momentum, which make up a tiny fraction of all excitons, recombine efficiently \cite{poellmann_resonant_2015}, while dark excitons are relatively stable. The fraction 
% of carriers that remains as an unbound electron-hole plasma recombines as well exhibiting the same spectral structure but different recombination rates. A very good review of this topic is given by Koch et al. 
% in \cite{koch_semiconductor_2006} and a more rigorous treatment is given in \cite{hoyer_influence_2003}. 
% The phonon-assisted relaxation of excitons has recently been discussed for TMDC semiconductors in \cite{selig_excitonic_2016}.

Experimental verification of the ionization equilibrium has been achieved in GaAs quantum wells using THz spectroscopy to probe transitions between 1s- and 2p-exciton states. \cite{kaindl_ultrafast_2003,koch_semiconductor_2006} 
A similar technique in the mid-infrared range has been applied recently to monolayer WSe$_2$. \cite{steinleitner_direct_2017}
Alternatively, the fractions of excitons and plasma can be determined from their contributions to photoluminescence (PL) spectra \cite{chatterjee_excitonic_2004} 
in combination with additional PL simulations.
As we suggest below, other ways to quantify the degree of exciton formation are angular-resolved photoemission spectroscopy (ARPES) and 
scanning tunneling spectroscopy (STS).
% to assign the contributions correctly as correlated and uncorrelated electron-hole pairs emit at the same energies.

In this paper, we combine calculations of material-realistic band structures and Coulomb matrix elements of the monolayer TMDC materials MX$_2$ (M$=$W,Mo and X$=$S,Se) with the state-of-the-art theory of ionization equilibrium, which we briefly introduce in the following section before discussing the results in detail. The theory is based on GW- and T-matrix self-energies describing the excited carriers and the effect of frequency-dependent screening. We also include excitonic screening that we find to be relevant and that is usually not taken into account. On these grounds, we study the influence of experimental and device-relevant parameters like dielectric screening, temperature and carrier doping on the ionization equilibrium and the Mott density.

%%%%%%%%%%%%%%%%%%%%%%%%%%%%%%%%%%%%%%%%%%%%%%%%%%%%%%%%%%%%%%%%%%%%%
%% THEORY PART
%%%%%%%%%%%%%%%%%%%%%%%%%%%%%%%%%%%%%%%%%%%%%%%%%%%%%%%%%%%%%%%%%%%%%

% \section{Ionization Equilibrium: Physical And Chemical Picture.}
\section{Spectral Functions and Exciton Satellites.}

To examine the equilibrium properties of excited carriers in TMDCs, we follow the approach developed in 
Refs.~\citenum{kremp_equation_1984,kremp_ladder_1984,zimmermann_mass_1985,kremp_quantum_1993} and reviewed in Ref.~\citenum{semkat_ionization_2009}. 
We use the quantum-statistical expression for the carrier density $n_a$ of the species $a$, which can be electrons or holes, as a function of temperature $T$ 
and chemical potential $\mu_a$ as a starting point:
\begin{equation}
  n_a(\mu_a,T)=\frac{i\hbar}{\mathcal{A}}\int_{-\infty}^{\infty}\frac{d\omega}{2\pi}\sum_{\bk\sigma}f^a(\omega)A_{\bk\sigma}^a(\omega)
     ~.
    \label{eq:carr_dens}
\end{equation}
$f^a(\omega)$ denotes the Fermi distribution function depending on $\mu_a$ and $T$, $\mathcal{A}$ is the crystal area and 
$A_{\bk\sigma}^a(\omega)=2i\textrm{Im}G^{\textrm{ret},a}_{\bk\sigma}(\omega)$ is the spectral function of the single-particle 
state $\ket{\bk\sigma a}$ related to the retarded single-particle Green's function
\begin{equation}
  G^{\textrm{ret},a}_{\bk\sigma}(\omega)=\frac{1}{\hbar\omega-\varepsilon_{\bk\sigma}^a-\Sigma_{\bk\sigma}^{\textrm{ret},a}(\omega)}
     ~.
    \label{eq:G_ret}
\end{equation}
The self-energy $\Sigma_{\bk\sigma}^{\textrm{ret},a}(\omega)$ accounts for many-particle effects giving rise to renormalizations of
the single-particle band structure $\varepsilon_{\bk\sigma}^a$ as well as contributions of bound states. For a given self-energy,
the inversion of Eq.~(\ref{eq:carr_dens}) yields the chemical potential $\mu_a(n_a,T)$ for each species and therefore any thermodynamic property 
of the system in the grand canonical description.
\begin{figure}[h!t]
\begin{center}
\includegraphics[width=1.\columnwidth]{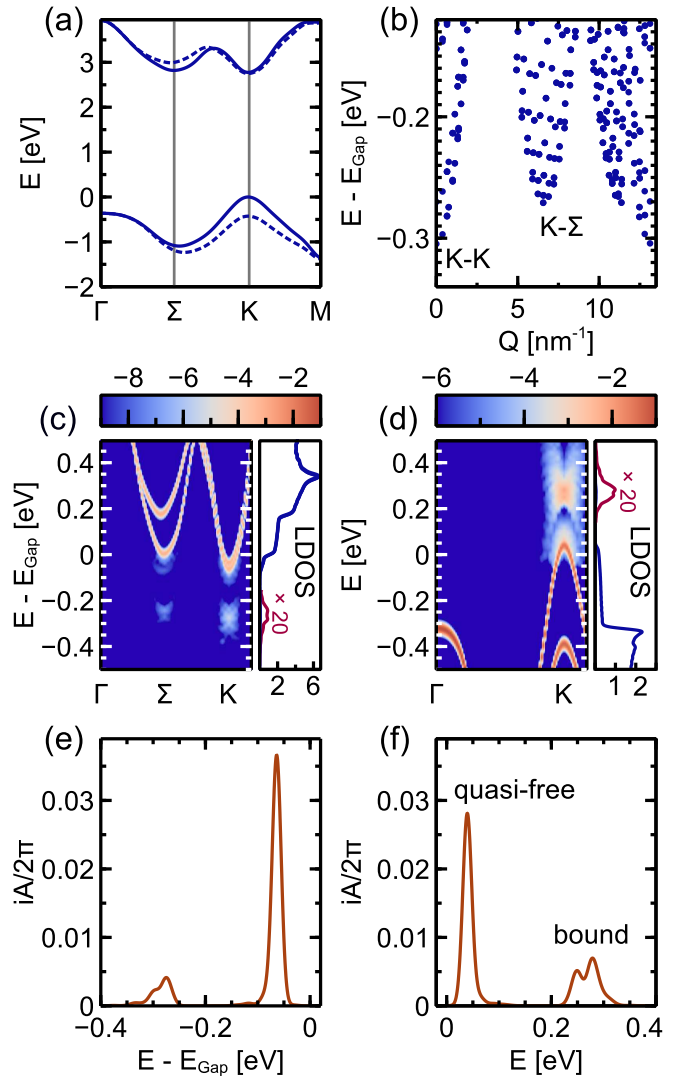}
\caption{
% \textbf{Spectral properties of the correlated electron-hole plasma in monolayer WS$_2$.} 
\textbf{Spectral properties of excited carriers in monolayer WS$_2$.} 
\textbf{(a)} Band structure of freestanding WS$_2$ as obtained 
from a G$_0$W$_0$ calculation at zero excitation density.
\textbf{(b)} Bound-state spectrum for WS$_2$ on SiO$_2$ substrate relative to the quasi-particle gap at zero excitation density
over the modulus of total exciton momentum $Q$ as obtained from a Bethe-Salpeter equation, see Eq.~(\ref{eq:bse}) in the Methods section.
Excitons involving the $\Gamma$, $K'$ and $\Sigma'$ valleys are included but not marked explicitely.
\textbf{(c)} Electron spectral function in extended quasi-particle approximation at $T=300$ K and excitation density 
$n_a=3.2\times10^{12}$~cm$^{-2}$ showing resonances from bound and quasi-free particles
in momentum-resolved representation on a logarithmic scale and as normalized local density of states (LDOS). 
Energies are measured relative to the quasi-particle band gap at zero excitation density. 
A phenomenological Gaussian broadening of $10$ meV (HWHM) is applied. 
\textbf{(d)} Hole spectral function in extended quasi-particle approximation.
\textbf{(e)} Electron spectral function as in (c) for spin-down electrons at the $K$-point. 
The vertical line marks the electron chemical potential. 
\textbf{(f)} Hole spectral function as in (d) for spin-up holes at the $K$-point.
}
\label{fig:spectralf}
\end{center}
\end{figure}
\begin{figure*}[h!t]
\begin{center}
\includegraphics[width=\textwidth]{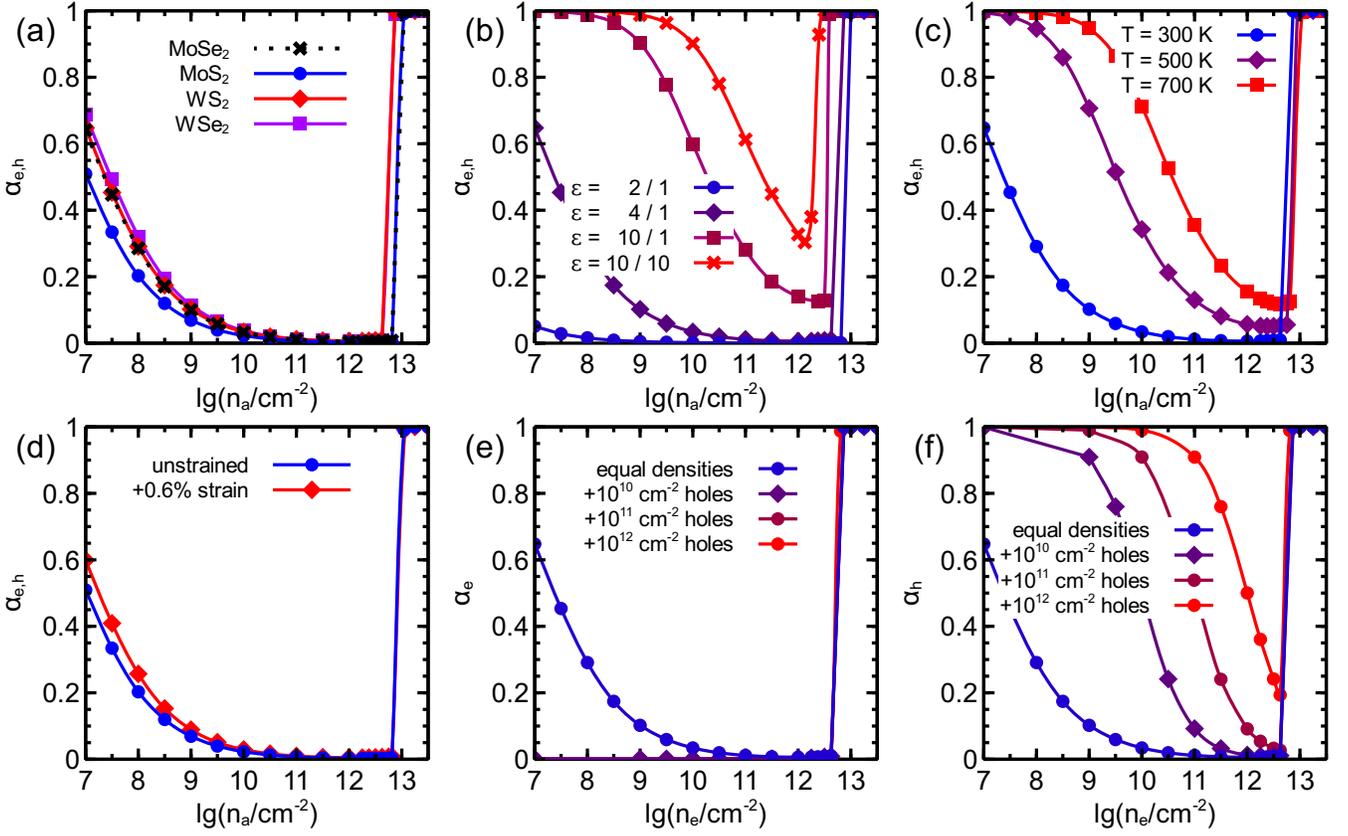}
\caption{\textbf{Degree of ionization as a function of excitation density} $n_a=n_{\textrm{free}}^{a}+n_X$, where $a$ denotes electrons or holes. In general, a fully ionized
plasma ($\alpha_a=1$) is found above the Mott density, excitons dominate ($\alpha_a\ll1$) below the Mott density and ionization appears again at low densities.  
\textbf{(a)} Comparison of different TMDC materials on SiO$_2$ substrate.
\textbf{(b)} Comparison of WS$_2$ in different dielectric environments with dielectric constant $\varepsilon$ of bottom/top environment.
\textbf{(c)} Comparison of WS$_2$ on SiO$_2$ substrate for different temperatures.
\textbf{(d)} Comparison of MoS$_2$ with different lattice constants on SiO$_2$ substrate. The unstrained layer corresponds to $a=3.18$~\AA\, 
while tensile biaxial strain is simulated by $a=3.20$~\AA\,.
\textbf{(e)} Fraction of ionized electrons for different levels of hole doping in WS$_2$ on SiO$_2$ substrate.
\textbf{(f)} Fraction of ionized holes for different levels of hole doping in WS$_2$ on SiO$_2$ substrate.}
\label{fig:ionization}
\end{center}
\end{figure*}
As we describe in detail in the Methods section, by using a T-matrix self-energy in screened ladder approximation and assuming small quasi-particle damping, 
we obtain a spectral function $A^a(\omega)$ in the so-called extended quasi-particle approximation. It exhibits poles for quasi-free and bound carriers 
as shown in Fig.~\ref{fig:spectralf}. 
From the multiple valleys in the single-particle band structure of electrons and holes (Fig.~\ref{fig:spectralf}(a)), a rich spectrum of bound states emerges (Fig.~\ref{fig:spectralf}(b)),
that contains a variety of dark excitons with large total momentum $Q$ besides the bright $K$-valley excitons commonly referred to as A and B. The dark excitons, though playing a minor role in optical 
experiments, are essential to the description of the ionization equilibrium.
% Beyond the valleys at the $K(K')$ point in reciprocal space hosting the prominent bright excitons, TMDC semiconductors exhibit additional valleys in the band structure at the $\Gamma$ point and at the 
% $\Sigma (\Sigma')$ point halfway between $K(K')$ and $\Gamma$ that allow for the formation of dark excitons with resonances close to the bright excitons, see Fig.~\ref{fig:spectralf}.
Various bound states are reflected in the low-energy satellites of the single-particle spectral function. 
%All of these states have to be taken into account when calculating the ionization equilibrium.
This shows that excitonic contributions are expected to be observed in experiments that are sensitive to these spectral properties. 
In ARPES \cite{damascelli_angle-resolved_2003},
momentum-resolved images of the electron spectral function comparable to Fig.~\ref{fig:spectralf}(c) are
obtained, weighted with Fermi distribution functions that are defined by the chemical potential $\mu_e$ and temperature $T$. For a fixed quasi-momentum state as shown in
Fig.~\ref{fig:spectralf}(e), this is typically referred to as energy distribution curve.
On the other hand, STS \cite{fischer_scanning_2007} probes the local density of states (LDOS) and thus momentum-averaged spectral functions of electrons and holes, which are
displayed in Figs.~\ref{fig:spectralf}(c) and (d).
% We therefore suggest that these experimental techniques are used to quantify the degree of exciton ionization and verify the results we present in this paper. 
We therefore propose to use these experimental techniques to spectrally distinguish between excitons and quasi-free carriers. This opens the possibility to 
quantify the degree of exciton ionization and verify the results we present in this paper.

\section{Degree of Ionization and Mott Transition.}

According to Eq.~(\ref{eq:carr_dens}), the spectral function can be used to separate the total electron and hole density ($a=e,h$), 
\begin{equation}
\begin{split} 
  n_a=n_{\textrm{free}}^a+n_X
     \,,
    \label{eq:density_final_short}
\end{split}
\end{equation}
into contributions from quasi-free carriers and from carriers bound as excitons, where the excitons are approximately treated as bosons. 
% Eq.~(\ref{eq:density_final_short}) corresponds to Eq.~(\ref{eq:density_final}) in the Methods section.
Hence the properties of the excited semiconductor at a given temperature and excitation density are defined by the density of electrons $n_e$, 
the density of holes $n_h$ and the density of excitons $n_X$. 
A certain degree of ionization of the excited carriers
\begin{equation}
\alpha_a=\frac{n_{\textrm{free}}^{a}}{n_a}
    \label{eq:ionization}
\end{equation}
will be established that is determined by the ionization equilibrium between electrons, holes and excitons. While for optical excitation, equal densities
of electrons and holes are generated, we distinguish here between electron and hole ionization 
to also include the effect of carrier doping where electron and hole densities are different.

Using single-particle band structures and bound-state spectra, which are determined as discussed in the Methods section, we solve Eq.~(\ref{eq:density_final_short}) numerically
to obtain the degree of ionization $\alpha_a$ in various TMDC materials under different experimental conditions.
% With the theory of ionization equilibrium outlined above and the bound-state spectra at hand, we can now discuss
% turn to the main subject of this Paper, which is 
% the balance between plasma and excitons as given by the 
% degree of ionization $\alpha_a$ defined in Eq.~(\ref{eq:ionization}) in various TMDC materials under different experimental conditions.
% It is obtained as root of the implicit equation (\ref{eq:density_final_short}),
% %
% \begin{equation}
% \begin{split} 
%   n_a=\alpha_a n_a+n_X[\alpha_e,\alpha_h]\,,
%     \label{eq:density_final_implicit}
% \end{split}
% \end{equation}
% %
% that corresponds to Eq.~(\ref{eq:density_final}) in the Methods section. The bound-carrier density $n_X$ depends on the degree of ionization via the electron and hole chemical 
% potentials and the quasi-particle renormalizations given 
% in Eq.~(\ref{eq:GW_energy}) that are treated self-consistently. 
The results are collected in Fig.~\ref{fig:ionization} and exhibit the behaviour of ionization degree as a function of excitation density.
% that has been discussed for conventional semiconductors \cite{semkat_ionization_2009}. 
There are different regimes of ionization to be observed. At high excitation densities between 
$3\times10^{12}$~cm$^{-2}$ and $1\times10^{13}$~cm$^{-2}$, depending on experimental parameters, efficient screening and
many-particle renormalizations lead to a full ionization of excited carriers, which is known as Mott effect. At lower densities around $n_a=1\times10^{12}$~cm$^{-2}$,
excitons dominate the physical properties of TMDCs for the parameters studied here due to the large exciton binding energies and density of states that
is mostly given by dark excitons. Bright excitons with very small momenta that are optically active make up only a tiny fraction of the total exciton density. 
As Fig.~\ref{fig:ionization}(b) shows, an efficient tuning knob for the degree of ionization is the
% Much stronger is the influence of
dielectric screening by the environment, which can change over a wide range depending on the experimental situation or device realization in which the TMDC monolayer is the active material. 
The reason is the strong impact of dielectric screening on the exciton binding energies.
Typical examples for substrates are Borofloat~($\varepsilon=2$), SiO$_2$~($\varepsilon=3.9$) and sapphire~($\varepsilon=10$). 
The dielectric constant of the environment on top of the monolayer is often given by the vacuum value. On the other hand, in devices the TMDC 
monolayer is fully encapsulated by dielectric material. As an example we consider a full dielectric environment with $\varepsilon=10$, which might be either sapphire or additional layers of TMDC material
in a vertical heterostructure whose main influence on the excitons is the dielectric screening. \cite{andersen_dielectric_2015} We find that the minimal degree of ionization can be tuned from below $0.1\%$
($99.9\%$ excitons) at weak dielectric screening to about $30\%$ at strong screening, while the Mott density is lowered at the same time by roughly a factor of $3$. 
The second important parameter that is relevant to applications of TMDC monolayers is the doping with additional carriers which might be either intrinsic or induced by external electric fields in a capacitor
structure. Here the fractions of ionized electrons and holes, $\alpha_e$ and $\alpha_h$, are discussed separately as the densities of the species are not equal anymore. We consider hole
doping of WS$_2$, but similar results are expected in case of electron doping. Looking at Figs.~\ref{fig:ionization} (e) and (f), we find that even at weak doping the minority
carriers are practically all bound as excitons below the Mott transition. On the other hand, at higher doping levels an increasing fraction of majority carriers exists as quasi-free plasma due to
missing partners for exciton formation. 
% The net effect is that in terms of 
% minority carriers, which limit the achievable exciton density, the Mott transition is lowered by roughly the density of doped excess carriers. 
As a function of minority-carrier density the Mott transition is lowered by roughly the density of doped excess carriers. 
From this we conclude that at doping levels above
$10^{13}$~cm$^{-2}$ neither dark nor bright excitons will exist in any case. This is supported by the experimental estimate for doping-induced 
ionization at several $10^{13}$~cm$^{-2}$. \cite{chernikov_electrical_2015}
Another crucial parameter is the temperature, see Fig.~\ref{fig:ionization}(c), which can vary in experiments or devices due to heating of the active material under strong optical or electrical pumping. This
effect has been taken into account to explain the observed exciton-to-plasma ratio in monolayer WSe$_2$ in Ref.~\citenum{steinleitner_direct_2017}.
At room temperature and even at elevated temperatures up to $700$ K excitons clearly dominate below the 
Mott transition. At the same time, the Mott density slightly increases with temperature due to weaker renormalizations of the quasi-particle gap. It turns out that strain
is no efficient tuning knob as both bright and dark excitons contribute to the ionization equilibrium, although bright excitons are preferred in moderately tensile-strained TMDCs, see the discussion in the Methods
section.
A comparison of different TMDC materials shows that excitons are slightly more important in molybdenum- than in tungsten-based TMDCs due to the larger binding energies, which leads to larger Mott densities.
\begin{figure*}[h!t]
\begin{center}
\includegraphics[width=\textwidth]{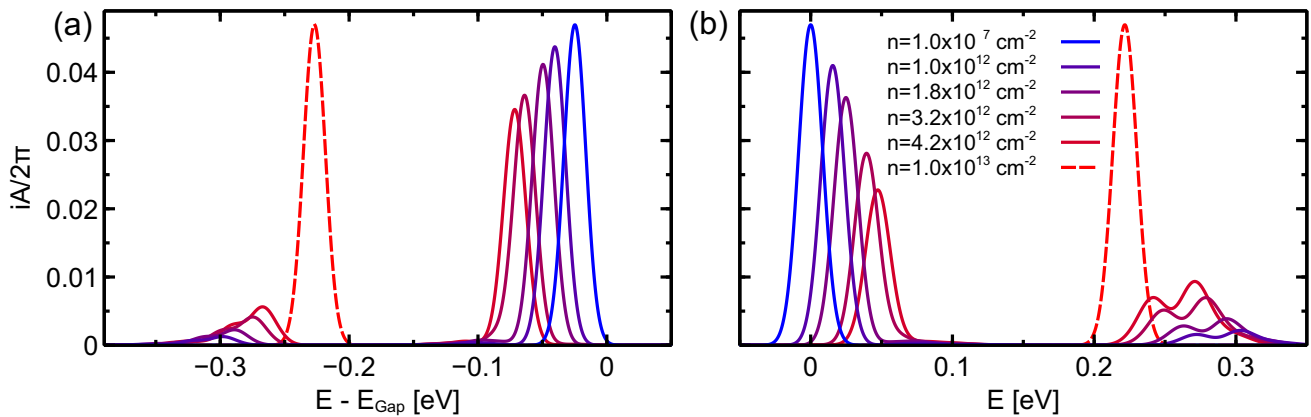}
\caption{\textbf{Excitation-density dependence of spectral functions.} 
Shown are spectral functions in extended quasi-particle approximation for spin-down electrons \textbf{(a)} and spin-up holes \textbf{(b)} at the $K$-point in WS$_2$ on SiO$_2$ substrate at $T=300$ K. 
A phenomenological Gaussian broadening of $10$ meV (HWHM) is applied. Electron energies are measured relative to the quasi-particle band gap at zero excitation density. Hole energies are shown as valence-band energies
to match Fig.~\ref{fig:spectralf}(f).}
\label{fig:spectralf_density}
\end{center}
\end{figure*}
%
% all materials are dominated by excitons up to fractions of $99.9\%$

\begin{figure}[h!t]
\begin{center}
\includegraphics[width=1.\columnwidth]{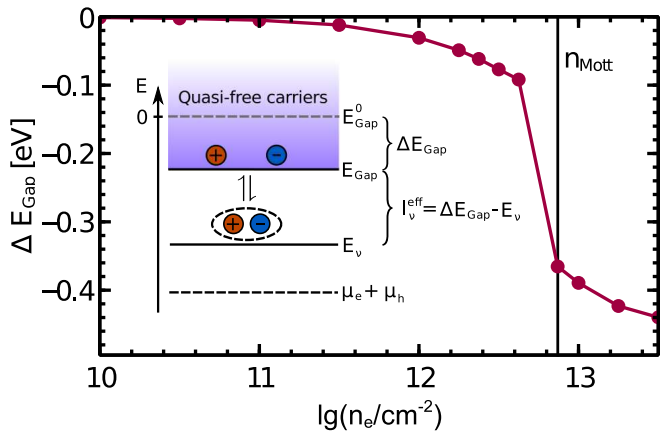}
\caption{\textbf{Ionization equilibrium and quasi-particle band-gap renormalization.} Shown is the band-gap renormalization $\Delta E_{\textrm{Gap}}$ for spin-up carriers at the $K$-point 
in WS$_2$ on SiO$_2$ substrate at $T=300$ K as function of excitation density. 
The vertical black line marks the Mott transition. The inset shows a schematic of the ionization equilibrium introducing the quasi-particle band gap at zero excitation density $E^0_{\textrm{Gap}}$,
which is conventionally set to zero energy. All quantities including the bound-state energies $E_{\nu}$ and the sum of electron and hole chemical potentials are measured relative to $E^0_{\textrm{Gap}}$.}
\label{fig:E_Gap}
\end{center}
\end{figure}

When approaching the Mott density from the low-density side, many-particle renormalizations become increasingly important, cf. Eq.~(\ref{eq:GW_energy}) in the Methods section.
% and drive the TMDCs towards an ionized state at even higher densities. 
Exchange interaction and efficient screening due to free carriers as well as excitons reduce the quasi-particle band gap and the exciton binding energies.
% as entering Saha's equation~(\ref{eq:saha}).
%As the excitons are assumed to be spectrally fixed, this corresponds to a reduction of exciton binding energy at the same time. 
More and more excitons are ionized, which leads to an increase of efficient free-carrier screening and thereby to a self-amplification of the ionization effect until 
all excitons are dissociated into an electron-hole plasma and the degree of ionization becomes $\alpha_a=1$. 
% This so-called Mott transition between excitons and electron-hole plasma,
% which we define as the degree of ionization $\alpha\approx 1$, appears at densities between $3\times10^{12}$~cm$^{-2}$ and $1\times10^{13}$~cm$^{-2}$ depending on experimental parameters. 
Note that $\alpha_a$ includes not only bright but also dark excitons with large total momentum for example between $K$ and $\Sigma$ valleys. 
They may have larger binding energies and be slightly more stable against ionization than bright excitons visible in an optical experiment.
Fig.~\ref{fig:spectralf_density} shows an illustration of the Mott effect in terms of the spectral functions in extended quasi-particle approximation, which
contain both exciton and quasi-free-particle signatures. At low excitation densities, the only spectral contribution stems from quasi-free carriers at the band edge. 
With increasing density, the quasi-particle peak is shifted to lower energies due to many-particle renormalizations. At the same time, spectral weight is transferred from the quasi-particle to the 
bound-state peaks as exciton populations increase, see the explicit expression of the spectral function in Eq.~(\ref{eq:spectral_fct_extended_T_matrix}). 
% The holes at the $K$-point form different excitons with electrons in $K$ and $\Sigma$ valleys leading to several peaks 
% below the quasi-particle gap while the electrons mainly participate in excitons involving $K$ and $K'$ valleys. 
The appearence of several exciton satellites in the hole spectral function is due to different bound states involving electrons either in the $K$- or $\Sigma$-valleys, see Fig.~\ref{fig:spectralf}~(b).
The spectral position of a bound-state peak in the spectral function of carrier $a$
is given by the difference of the corresponding exciton energy $E^{ab}$ and the energy of the second carrier $b$ involved in the bound state. 
The bound resonance might therefore be interpreted as an effective ionization energy of the actual carrier $a$ with respect to its energy in the quasi-particle band structure. 
This underlines the fact that excitonic signatures are observable in experiments that are sensitive to the single-particle spectral function, such as ARPES and STS. Note that while the amplitudes of
bound-state resonances in the spectral functions are relatively small, observables like the carrier density (\ref{eq:carr_dens}) and the photoemission intensity 
involve weighting with a Fermi function that strongly favors the low-energy
resonances over the quasi-free contribution. With increasing excitation density, quasi-particle and excitonic resonances approach until above the Mott density all excitons are ionized and only a quasi-particle peak remains.
Fig.~\ref{fig:E_Gap} shows the reduction of the quasi-particle gap until the Mott transition appears around $n_a=8\times10^{12}$~cm$^{-2}$.
%
% With increasing excitation density, excitons begin to dominate the physical properties of TMDCs for the parameters studied here due to the large exciton binding energies and density of states that
% is mostly given by dark excitons. Bright excitons with very small momenta that are optically active make up only a tiny fraction of the total exciton density. Around $n_a=1\times10^{12}$~cm$^{-2}$ 
% all materials are dominated by excitons up to fractions of $99.9\%$ before many-particle renormalizations become important and drive the TMDCs towards an ionized state at even higher densities.
% Moreover, the Mott densities given here should be understood as 
% approximate values due to the limitations of the extended quasi-particle approximation.
% At $n\approx 4\times10^{12}$~cm$^{-2}$, it exceeds the binding energy of the A exciton suggesting that the latter 
% dissociates already around this density. 
% The total Mott transition given by $\alpha\approx 1$ appears around $n=8\times10^{12}$~cm$^{-2}$.

An alternative picture of the interacting electrons and holes, that is consistent with the extended quasi-particle approximation, is the so-called \textit{chemical picture}
in which excitons are considered as a new particle species besides electrons and holes. \cite{kremp_quantum_2005,semkat_ionization_2009} They are characterized 
by a chemical potential 
\begin{equation}
\mu_{X,\nu}=\mu_e+\mu_h-E_{\nu}\,,
    \label{eq:gen_MAL}
\end{equation}
with bound-state energies $E_{\nu}$ that are given by the relative motion of electron and hole, and an ideal Bose distribution function. 
In the chemical picture, solving Eq.~(\ref{eq:density_final_short}) corresponds to an adaption of the chemical potentials of the different particle species, 
namely electrons, holes and excitons, as in a chemical reaction.
These considerations are consistent with the theory based on spectral functions, that 
we use to obtain all numerical results presented in this paper. Only for the purpose of illustration, we simplify the theory considering
the nondegenerate case ($f^a(E^a_{\bk\sigma})\ll 1$) and a single band-structure valley for electrons and holes each.
% approximately split the chemical potentials
% of electrons and holes into an ideal free-carrier 
% and a correlation contribution: $\mu_a=\mu_a^{\textrm{id}}+\Delta\mu_a$.
% The latter is given by the many-particle renormalizations of electron and hole energies, see Eq.~(\ref{eq:GW_energy}) in the Methods section, and lowers the chemical potentials. 
% In the nondegenerate case ($f^a(e^a_{\bk\sigma})\ll 1$)
% and for a single band-structure valley for electrons and holes each, 
Then a Saha equation can be formulated that determines the degree of ionization:
% %
% \begin{equation}
% \frac{1-\alpha}{\alpha^2}\propto n_e e^{-\beta(E_{\nu}-\Delta\mu_e-\Delta\mu_h)} \,.
%     \label{eq:saha}
% \end{equation}
% %
%
\begin{equation}
\frac{n_X}{n_e n_h}\propto e^{\beta I_{\nu}^{\textrm{eff}}} \,.
    \label{eq:saha}
\end{equation}
In analogy to the usual mass action law, $I_{\nu}^{\textrm{eff}}=\Delta E_{\textrm{Gap}}-E_{\nu}$ 
% obeying the generalized mass action law expressed by Eq.~(\ref{eq:gen_MAL}). 
%with the excitation-induced shift $\Delta E_{\nu}$ of the exciton energy 
can be interpreted as an effective ionization potential of excitons that corresponds to the exciton binding energy, see also the inset in Fig.~\ref{fig:E_Gap}. 
It is obvious from Saha's equation, that a large exciton binding energy and low temperature favor the formation of excitons versus the dissociation into an unbound electron-hole plasma.
The ionization potential depends on excitation density as a consequence of the excitation-induced
lowering of the band continuum edge given by $\Delta E_{\textrm{Gap}}$ and the shift of the bound-state energy $E_{\nu}$. 
The bound-state shift on the other hand is a net result of band-gap shrinkage, screening of exciton binding energy and Pauli blocking \cite{steinhoff_influence_2014} and is much weaker than 
the band-gap shift due to compensation effects. 
In the end, the ionization potential is lowered with increasing excitation density until at $I_{\nu}^{\textrm{eff}}=0$
the bound state vanishes and merges with the continuum edge, which is the Mott effect. 

A striking observation in Fig.~\ref{fig:ionization} is the degree of ionization approaching unity at low excitation densities, which is somewhat counter-intuitive but can be 
understood from a thermodynamical point of view. The potential
that is minimized by the many-particle system is the free energy $F=U-TS$. At low densities and fixed temperature, the entropy $S$ gained by a dissociation of an exciton into two separate particles
overcompensates the reduction of internal energy $U$ by the exciton binding energy $E_B$. Hence the so-called entropy ionization already discussed by Mock et al.~\cite{mock_entropy_1978} is connected 
to the huge phase space available for quasi-free carriers in the low-density limit.
We may clarify this using the entropy of an ideal gas with $N$ particles in a volume $V$ as given by the Sackur-Tetrode equation:
\begin{equation}
\begin{split} 
  S=N k_B\left[\textrm{ln}\left(\frac{V}{N} c(T)\right) +\frac{5}{2}\right]
     \,,
    \label{eq:entropy_ideal_gas}
\end{split}
\end{equation}
where $c(T)$ is a temperature-dependent parameter. Obviously, the dissociation of an exciton gas ($N$ particles) into a free electron-hole plasma ($2N$ particles) yields the entropy 
$\Delta S=N k_B \textrm{ln}\left(n^{-1}\right)$ with $n=N/V$ up to some additive constant. It follows that the critical density $n_{\textrm{crit}}$ below which the free energy is dominated by entropy essentially 
scales as $\textrm{exp}\left(-E_B/k_B T\right) $ with temperature. 
% The so-called entropy ionization has already been discussed by Mock et al. \cite{mock_entropy_1978} 
% It implies that a single exciton that is optically generated in a macroscopic crystal would thermally dissociate if the recombination time is long enough.
% Looking at Fig.~\ref{fig:ionization}(e), we find that even at weak doping the minority
% carriers are practically all bound as excitons below the Mott transition, as the total (minority plus majority) carrier density is increased above the critical density $n_{\textrm{crit}}$ that marks the entropy-dominated regime.

Although the extended quasi-particle approximation and the chemical picture are very descriptive, we have to be aware of their limitations. 
The approach relies on the assumption of a quasi-equilibrium of both types of carriers,
which tends to overestimate the fraction of bound carriers as exciton formation takes a certain time after excitation of electron-hole pairs.
Nevertheless, we expect the approach to give a reasonable quantitative description of the ionization equilibrium, particularly of the trends that can be expected under variation of external
experimental parameters.
% and a more sophisticated theory is required. 
A rather fundamental discussion is concerned with the Mott transition as a first-order phase transition between an exciton gas and a fully ionized electron-hole plasma. \cite{kremp_quantum_2005, semkat_ionization_2009} 
The phase transition would be connected to an instability of thermodynamic functions that manifests itself in an ambiguity of $\alpha_a$ in a certain region below the Mott density. 
Due to excitation-induced broadening of the two-particle states, which is assumed small in our approach, and the shrinkage of the ionization potential towards the Mott transition, 
quasi-free and bound carriers cannot really be separated in this density regime.
We avoid this regime as 
% the picture of separate particle species breaks down and 
a more sophisticated theory including full spectral
functions and exciton-exciton interaction would be required. Also, screening in a correlated many-particle system near the Mott transition is an intricate problem \cite{kraeft_quantum_1986}.
% In our results we find no
% indication of such an ambiguity, the transition from excitons to plasma being smooth for all parameters studied here. It is possible that ambiguities would appear for example at lower temperatures but even
% then it is unclear if the instability is rather an artifact of the theory than real physics. \rem{kremp 2005}
However, we observe that taking into account excitonic screening is necessary to obtain meaningful results around the minima of $\alpha_a$ in the exciton-dominated regime. 
% Otherwise, the many-particle renormalizations (\ref{eq:GW_energy}) that reduce the ionization potentials
% of excitons would not scale with the total density of carriers but only with the fraction $\alpha$. 
Otherwise, coming from the low-density side of the ionization curves in Fig.~\ref{fig:ionization}, there would be no mechanism to efficiently break up the excitonic binding 
and trigger the transition to an ionized plasma. We discuss the contribution of excitons to screening in the following section.

%As carrier-phonon interaction depends only weakly on the carrier density, which is the main parameter that
%drives the ionization equilibrium, we assume that its influence on the results is negligible, although it is essential to establish the quasi-equilibrium in the first place. Its main effects would be a polaron
%shift that adds to the ab initio band structure and some additional broadening of exciton lines which is for most parts of the carrier-density range small compared to the exciton binding energies.

Another prominent feature of TMDC semiconductors is the formation of trions, which could in principle be included as additional particle species in the spirit of the chemical picture. \cite{kremp_quantum_1993}
In practice, obtaining bound-trion spectra on the same footing as excitons is a very challenging task of its own that is beyond the scope of this paper. Due to the relatively small trion binding energies, we
assume that their influence on our results are negligible.

% Having introduced the physical and chemical pictures of exciton ionization, the next section is devoted to a detailed discussion of bound-state spectra as essential ingredients 
% of the theory before we turn to numerical results for the ionization equilibrium itself.

% the calculation of bound-state energies and wave functions
% that enter the theory as an essential ingredient via the T-matrix.

\section{Excitonic Screening.}

In the spirit of the extended quasi-particle approximation to the spectral function, there are two types of contributions to excited-carrier screening of the Coulomb interaction, 
the metal-like free-carrier screening and dipolar screening due to bound excitons. %, see Eq.~(\ref{eq:screening}).
\begin{figure}[h!t]
\begin{center}
\includegraphics[width=\columnwidth]{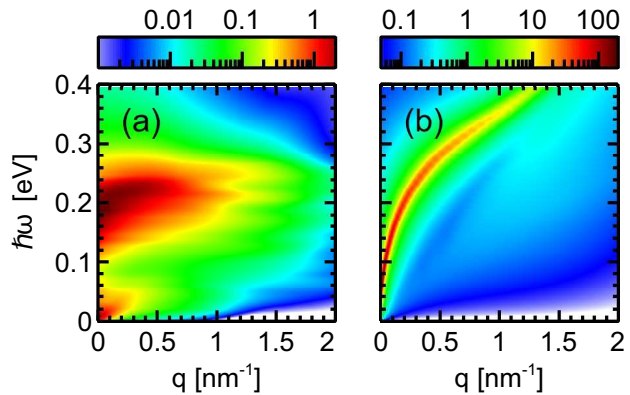}
\caption{\textbf{Characterization of excited-carrier screening.} 
Shown is the plasmon spectral function given by the imaginary part of the inverse dielectric function including the 2-d-Coulomb singularity $\varepsilon^{-1,\textrm{ret}}_{\bq}(\omega)/|\bq|$ for momenta
along the contour $\Gamma$-$K$ in WS$_2$ on SiO$_2$ substrate at $T=300$ K. A quasi-particle broadening of $10$ meV is used. The carrier densities are 
 \textbf{(a)} $n_{\textrm{free}}=2.7\times10^{10}$ cm$^{-2}$, $n_X=3.1\times10^{12}$ cm$^{-2}$ and \textbf{(b)} $n_{\textrm{free}}=1.8\times10^{13}$ cm$^{-2}$, $n_X=0$ corresponding to
points on the ionization equilibrium curve in Fig.~\ref{fig:ionization} in the exciton-dominated and the plasma-dominated regime, respectively.}
\label{fig:plasmon}
\end{center}
\end{figure}
%
% Excitonic screening is an interesting topic of its own that has not been discussed much in literature. 
The screening can be characterized by the plasmon spectral function, see Eq.~(\ref{eq:W_spec}), that contains excitations in the interacting electron-hole plasma as poles in the 
$q$-$\omega$-plane, see Fig.~\ref{fig:plasmon}. In the exciton-dominated regime shown in Fig.~\ref{fig:plasmon}(a), besides the usual 2-d free-carrier contribution at small energies and small momenta 
a broad resonance above $150$ meV appears. It stems from transitions between 1s- and 2s-like exciton states, see Fig.~\ref{fig:spectralf}(b), 
and also from comparable transitions between exciton states with large momenta.
% The selected results presented in Fig.~\ref{fig:plasmon} reveal distinct features of excitonic screening  
There are contributions at smaller energies as well that can not be as easily distinguished from free-carrier screening.
At large densities beyond the Mott transition the plasmon spectral function shows a pronounced peak structure with a square-root-like behaviour at small momenta 
which has been discussed for TMDCs in Ref.~\citenum{groenewald_valley_2016} and which is typical for a two-dimensional electron gas. \cite{haug_quantum_1993}
Although excitons are expected to be much less polarizable than a free electron-hole plasma and hence contribute less to screening, at elevated excitation densities with a large 
fraction of carriers bound as excitons, their contribution can be essential. From the many-particle renormalization caused by free-carrier and excitonic screening, 
we deduce that in monolayer TMDC semiconductors excitonic screening is less efficient by two to three orders of magnitude for comparable excitation
densities. Nevertheless, in a regime where more than $99\%$ of carriers are bound as excitons, excitonic screening still yields a significant contribution. 
As the plasmon spectral function is directly observable by electron energy loss spectroscopy (EELS) \cite{kogar_exciton_2016}, we suggest to use this technique to explore exciton signatures in the dielectric function
eperimentally.
% Besides the twofold screening, the extended quasi-particle spectral function also takes care of the fact that exchange interaction, which lowers the single-particle
% energies, is not only sensitive to quasi-free carriers but also to carriers bound as excitons, see Eq.~(\ref{eq:HF_self_energy}).

% \rem{bright-exciton density?}
% We can also estimate the density of bright excitons, calculation procedure, fuer ein beispiel angeben (welcher anteil X ist hell?), bedenken, dass dies nicht direkt aussage ueber quantum yield ist
% wegen relaxation von Q>0 exzitonen und plasma emission (s koch)

%%%%%%%%%%%%%%%%%%%%%%%%%%%%%%%%%%%%%%%%%%%%%%%%%%%%%%%%%%%%%%%%%%%%%
%% CONCLUSION PART
%%%%%%%%%%%%%%%%%%%%%%%%%%%%%%%%%%%%%%%%%%%%%%%%%%%%%%%%%%%%%%%%%%%%%

\section{Conclusion.} 

The exciton ionization equilibrium in monolayer TMDC semiconductors has been studied for various material as well as experimentally and device-relevant external parameters 
on the basis of an ab initio description of the electronic band structure and Coulomb interaction. We observe entropy ionization of excitons at
low excitation densities 
% that are relevant for PL experiments 
and a Mott transition to a fully ionized plasma at high densities between $3\times10^{12}$~cm$^{-2}$ and $1\times10^{13}$~cm$^{-2}$ depending on 
experimental parameters. Below the Mott transition, excitons become dominant in all cases with maximal fractions of excitons between $70\%$ and more than $99.9\%$. The most efficient tuning knobs are dielectric screening of the Coulomb interaction via the choice of dielectric environment and carrier doping that can induce complete ionization above a level of $10^{13}$~cm$^{-2}$. We suggest that fingerprints of excitonic contributions 
can be observed in ARPES and STS experiments, which are sensitive to the single-particle spectral functions. Moreover, we find that excitonic screening, although two to
three orders of magnitude less efficient than free-carrier screening at comparable excitation densities, plays an important role in the description of ionization equilibrium. Exciton signatures in the
dielectric function suggest EELS as another way to study the ionization equilibrium in excited semiconductors.

\section{Methods.}

\subsection{Theory of Ionization Equilibrium.}

We start from the general expression for the carrier density (\ref{eq:carr_dens}) and the spectral function
\begin{equation}
  A_{\bk\sigma}^a(\omega)=2i\textrm{Im}\,\frac{1}{\hbar\omega-\varepsilon_{\bk\sigma}^a-\Sigma_{\bk\sigma}^{\textrm{ret},a}(\omega)}
     ~.
    \label{eq:G_spec}
\end{equation}
In the limit of small quasi-particle damping $\textrm{Im}\,\Sigma^{\textrm{ret},a}\ll\textrm{Re}\,\Sigma^{\textrm{ret},a}$, the spectral function can
be expanded in linear order of $\textrm{Im}\,\Sigma^{\textrm{ret},a}$ yielding the carrier density in so-called extended quasi-particle approximation
\begin{equation}
\begin{split} 
  &n_a(\mu_a,T)\\ & =\frac{1}{\mathcal{A}}\sum_{\bk\sigma}f^a(E^a_{\bk\sigma}) \\ & -\frac{1}{\mathcal{A}}\sum_{\bk\sigma}\int_{-\infty}^{\infty}\frac{d\omega}{2\pi}
  \frac{2}{\hbar}\textrm{Im}\,\Sigma_{\bk\sigma}^{\textrm{ret},a}(\omega)\left[f^a(E^a_{\bk\sigma})-f^a(\omega)\right]\\
  &\frac{d}{d\omega}\frac{\mathcal{P}}{\omega-E^a_{\bk\sigma}/\hbar}\\
  &=n_a^{\textrm{QP}}+n_a^{\textrm{corr}}
  ~,
    \label{eq:n_ext_qp_approx}
\end{split}
\end{equation}
where the quasi-particle energy $E^a_{\bk\sigma}$ is given by $E^a_{\bk\sigma}=\varepsilon_{\bk\sigma}^a+\textrm{Re}\,\Sigma_{\bk\sigma}^{\textrm{ret},a}(E^a_{\bk\sigma})$ 
and $\mathcal{P}$ denotes the Cauchy principal value. \cite{kremp_equation_1984,semkat_ionization_2009} The total density is divided into 
contributions from quasi-free particles and correlated particles, the latter being either in bound or scattering many-particle states.

The spectral function in extended quasi-particle approximation corresponding to this separation into free and correlated carriers is given by
\begin{equation}
\begin{split} 
  A_{\bk\sigma}^a(\omega) = &-2\pi i\delta(\hbar\omega-E^a_{\bk\sigma})(1-Z_{\bk\sigma}^a) \\
  &- 2\pi i \Gamma_{\bk\sigma}^a(\omega)
  ~,
    \label{eq:spectral_fct_extended}
\end{split}
\end{equation}
with $ \Gamma_{\bk\sigma}^a(\omega) = \textrm{Im}\Sigma_{\bk\sigma}^{\textrm{ret},a}(\omega)\frac{1}{\pi}\frac{d}{d\hbar\omega}\frac{\mathcal{P}}{\hbar\omega-E^a_{\bk\sigma}} $ and
the renormalization factor $Z_{\bk\sigma}^a = \int d\hbar\omega \Gamma_{\bk\sigma}^a(\omega)$. The first term describes quasi-free particles at renormalized energies. Their spectral weight 
is reduced according to the renormalization factor to account for correlated carriers, which are spectrally described by the second term. 
% Examples for spectral functions are shown and discussed below.

To evaluate the expressions (\ref{eq:n_ext_qp_approx}) and (\ref{eq:spectral_fct_extended}), we have to choose an approximation for the self-energy 
$\Sigma^{\textrm{ret},a}(\omega)$. The real and imaginary parts of $\Sigma$ determine the quasi-particle energies and the correlated part of the carrier density, 
respectively. An appropriate choice is the screened ladder approximation \cite{stolz_correlated_1979,kremp_equation_1984,kremp_quantum_1993} 
$\Sigma(\omega)=\Sigma^{\textrm{H}}+\Sigma^{\textrm{GW}}(\omega)+\Sigma^{\textrm{T}}(\omega)$ that takes into account screening
of Coulomb interaction due to excited carriers as well as the formation of bound two-particle states and consists of Hartree, GW and T-matrix contributions.
We assume that renormalizations due to the Hartree self-energy are small compared to exchange and correlation effects.
In the T-matrix contribution, we neglect exchange terms and assume static screening so that the T-matrix depends only on one instead of three frequency arguments.
Thus we obtain for the imaginary part of the self-energy using the generalized Kadanoff-Baym ansatz \cite{kremp_quantum_2005}:
\begin{equation}
\begin{split} 
  &\textrm{Im}\,\Sigma_{\bk\sigma}^{\textrm{ret},a}(\omega) \\
  &=\frac{1}{\mathcal{A}}\sum_{\bk b}V^{ab}_{\bk\bk'\bk\bk'}\textrm{Im}\,\varepsilon^{-1,\textrm{ret}}_{\bk-\bk'}(\omega-E^b_{\bk'\sigma}/\hbar)\\
  &\left[f^b(E^b_{\bk'\sigma})-n^B(E^b_{\bk'\sigma}-\hbar\omega)\right]\\
  &+\frac{1}{\mathcal{A}}\sum_{\bk b\sigma'}\textrm{Im}\,T''^{,\textrm{ret},ab}_{\bk\bk'\sigma\sigma'}(\omega+E^b_{\bk'\sigma'}/\hbar)\\
  &\left[f^b(E^b_{\bk'\sigma'})+n^B_{ab}(\hbar\omega+E^b_{\bk'\sigma'})\right]
     ~.
    \label{eq:im_self_en_screened_ladder}
\end{split}
\end{equation}
Here we applied thermal equilibrium relations for the screened Coulomb interaction \cite{kremp_quantum_2005}:
\begin{equation}
\begin{split} 
  & V^{\textrm{S},<,ab}_{\bk\bk'\bk\bk'}(\omega)=n^B(\omega)V^{ab}_{\bk\bk'\bk\bk'}2i\textrm{Im}\,\varepsilon^{-1,\textrm{ret}}_{\bk-\bk'}(\omega), \\
  & V^{\textrm{S},>,ab}_{\bk\bk'\bk\bk'}(\omega)=(1+n^B(\omega))V^{ab}_{\bk\bk'\bk\bk'}2i\textrm{Im}\,\varepsilon^{-1,\textrm{ret}}_{\bk-\bk'}(\omega)\,.
    \label{eq:thermal_plasmon_prop}
\end{split}
\end{equation}
$\varepsilon^{-1,\textrm{ret}}_{\bq}(\omega)$ is the longitudinal dielectric function describing screening due to excited carriers and $n^B(\omega)$ is the Bose distribution function 
of the elementary plasma excitations called plasmons. $V^{ab}$ denotes Coulomb matrix elements between species $a$ and $b$ which contain dielectric screening due to carriers in the ground state and due to
the environment but no screening due to excited carriers. $T''$ denotes the T-matrix with the two lowest-order terms subtracted from the ladder expansion and is discussed in the following subsection.

\subsection{T-matrix and Bound Carriers.}

The T-matrix in statically screened ladder approximation describing bound and scattering two-particle states between carrier species $a$ and $b$ obeys a Lippmann-Schwinger equation (LSE)
\begin{equation}
  T^{\textrm{ret},ab}(\omega)=V^{\textrm{S},\textrm{ret},ab}+i\hbar V^{\textrm{S},\textrm{ret},ab}\,\mathcal{G}^{\textrm{ret},ab}(\omega)\, T^{\textrm{ret},ab}(\omega)
     \,,
    \label{eq:LSE}
\end{equation}
where $\mathcal{G}^{\textrm{ret},ab}(\omega)$ is the free two-particle Green's function in the particle-particle channel. The corresponding interacting
two-particle Green's function $G^{\textrm{ret},ab}_2(\omega)$ fulfills a Bethe-Salpeter equation, that has been discussed in detail in \cite{kraeft_quantum_1986,bornath_two-particle_1999}
and is equivalent to the LSE. We will exploit this fact later when solving the LSE and evaluating the T-matrix self-energy.
In its homogeneous form, the BSE in static ladder approximation is given by
\begin{equation}
\begin{split} 
&0=(E_{\bk\sigma}^a+E_{\bk+\bQ\sigma'}^b)G^{\textrm{ret},ab}_{2,\bk,\bk+\bQ\sigma\sigma'}  \\
  -&(1-f_{\bk\sigma}^a-f_{\bk+\bQ\sigma'}^b)\frac{1}{\mathcal{A}}\sum_{\bk'}V^{\textrm{S},ab}_{\bk+\bQ,\bk',\bk,\bk'+\bQ}G^{\textrm{ret},ab}_{2,\bk',\bk'+\bQ\sigma\sigma'} 
%   \\
%   =& E^{\sigma\sigma'}_{\nu\bQ}G^{\textrm{ret},ab}_{2,\bk,\bk+\bQ\sigma\sigma'}  
  \,.
    \label{eq:bse}
    \end{split} 
\end{equation}
Diagonalization yields bound states $\ket{\nu\sigma\sigma'\bQ}$ and eigenenergies $E^{\sigma\sigma'}_{\nu\bQ}$. We drop the indices $a$ and $b$ here, assuming that only two-particle states between 
different carrier species are involved. Due to the translational invariance of the crystal, the bound states can be classified by 
the total exciton momentum $\bQ$ as discussed in \cite{qiu_nonanalyticity_2015}. Here we neglect the effect of electron-hole exchange interaction that leads to a fine-structure 
splitting of excitons and trions \cite{qiu_nonanalyticity_2015,jones_excitonic_2016,plechinger_trion_2016} in the meV range, which is small compared to the exciton binding energies of several hundred meV. 
As a consequence, electron and hole spins, 
which are already good quantum numbers in monolayer TMDC materials due to crystal symmetry, also classify the bound states. For each total momentum and spin combination a series
of excitons exists, which is labeled here by $\nu$, analogue to the angular momentum states of Hydrogen-like Wannier excitons. Due to the two-dimensional nature of monolayer TMDCs and the related strong 
momentum dependence of dielectric screening, nontrivial exciton series deviating from a Hydrogen-like
spectrum are observed. \cite{qiu_optical_2013,chernikov_exciton_2014,berghauser_analytical_2014} The eigenenergies decompose into a part from the relative motion of 
electron and hole and a kinetic part depending on the total momentum: $E^{\sigma\sigma'}_{\nu\bQ}=E^{\sigma\sigma'}_{\textrm{rel},\nu}+E^{\sigma\sigma'}_{\textrm{kin},\bQ}$.
We can use Bloch basis functions to find a representation of the bound states corresponding to exciton wave functions
\begin{equation}
\begin{split} 
& \psi^{\sigma\sigma'}_{\nu\bQ}(\bk)= \big< \bk\bk'\sigma\sigma'ab \big| \nu\sigma\sigma'\bQ \big> \delta_{\bk',\bk+\bQ}\,,
    \label{eq:wave_functions}
    \end{split} 
\end{equation}
where $\bk$ conventionally denotes the hole momentum, while the electron momentum is fixed via the total momentum.

An explicit expression for the T-matrix can be obtained by writing the LSE (\ref{eq:LSE}) in the basis of two-particle eigenstates $\ket{\nu\sigma\sigma'\bQ}$ as shown in detail
in Ref.~\citenum{kremp_quantum_2005}. Since the BSE represents a generalized eigenvalue problem, the eigenstates form a biorthogonal basis. The procedure yields a spectral representation
of the T-matrix in operator form that is referred to as \textit{bilinear expansion}:
\begin{equation}
\begin{split} 
&T^{\textrm{ret},ab}(\omega)= \\
&\sum_{\nu\sigma\sigma'\bQ}N_{ab}^{-1}(E_{\textrm{kin}}-\hbar\omega)\frac{\ket{\nu\sigma\sigma'\bQ}\bra{\nu\widetilde{\sigma\sigma'}\bQ}}{\hbar\omega - E^{\sigma\sigma'}_{\nu\bQ}}(E_{\textrm{kin}}-E^{\sigma\sigma'}_{\nu\bQ})
    \label{eq:bilinear}
    \end{split} 
\end{equation}
with the Pauli blocking factor $N_{ab}=1-f^a-f^b$, the operator of kinetic energy $E_{\textrm{kin}}$ and the eigenstate of the adjoint BSE $\bra{\nu\widetilde{\sigma\sigma'}\bQ}$. 
The bilinear expansion is used in the following to evaluate the imaginary part of the self-energy (\ref{eq:im_self_en_screened_ladder}) and thereby the contribution of correlated carriers.

\subsection{Separation of Bound and Quasi-Free Carriers.}

Inserting Eq.~(\ref{eq:im_self_en_screened_ladder}) into Eq.~(\ref{eq:n_ext_qp_approx}) and noting that neither the GW self-energy nor the two lowest
T-matrix terms contribute to the carrier density \cite{stolz_correlated_1979}, we obtain \cite{kremp_quantum_1993,semkat_ionization_2009}
\begin{equation}
\begin{split} 
  n_a(\mu_a,T)&=\frac{1}{\mathcal{A}}\sum_{\bk\sigma}f^a(E^a_{\bk\sigma})\\
  &+\frac{1}{\mathcal{A}^2}\sum_{\bk\bk' b\sigma'}\int_{-\infty}^{E_{\textrm{Gap}}}\frac{d\omega}{\pi}n^B_{ab}(\omega)\\
  &\textrm{Im}
  \,T^{\textrm{ret},ab}_{\bk\bk'\sigma\sigma'}(\omega)\frac{d}{d\omega}i\hbar\mathcal{G}^{\textrm{ret},ab}_{\bk\bk'\sigma\sigma'}(\omega)\\
  &+n_{\textrm{scatt}}
     \,.
    \label{eq:density_full}
\end{split}
\end{equation}
$n^B_{ab}(\omega)=\big[\textrm{exp}(\beta(\hbar\omega-\mu_a-\mu_b))-1\big]^{-1}$ is the Bose distribution function depending on the chemical potentials of both carrier species.
Eq.~(\ref{eq:density_full}) contains contributions of both bound two-particle states
(below the single-particle gap $E_{\textrm{Gap}}$) and scattering two-particle states, the latter being explicitely given in Refs.~\citenum{kremp_ladder_1984,zimmermann_mass_1985,kremp_quantum_1993}. 
The renormalization factor $Z_{\bk\sigma}^a$ of the quasi-particle resonance in the spectral function (\ref{eq:spectral_fct_extended})
enters the contribution of correlated carriers as Pauli-blocking factor and as correction to the two-particle scattering spectrum.
To simplify the following discussion, we neglect the contribution $n_{\textrm{scatt}}$ of scattering states beyond the quasi-free carriers and consider only the bound-state contribution 
given by the real-frequency poles of the T-matrix \cite{kremp_ladder_1984}:
\begin{equation}
\begin{split} 
  &\frac{1}{\mathcal{A}^2}\sum_{\bk\bk' b\sigma'}\textrm{Im}
  \,T^{\textrm{ret},ab}_{\bk\bk'\sigma\sigma'}(\omega)\frac{d}{d\omega}i\hbar\mathcal{G}^{\textrm{ret},ab}_{\bk\bk'\sigma\sigma'}(\omega)\Big|_{\textrm{bound}} \\
   &=\pi\frac{1}{\mathcal{A}}\sum_{\nu\sigma'\bQ}\delta(\hbar\omega - E^{\sigma\sigma'}_{\nu\bQ})\,.
    \label{eq:dens_bound}
\end{split}
\end{equation}
Using Eq.~(\ref{eq:dens_bound}), we arrive at the final expression for the carrier density:
\begin{equation}
\begin{split} 
  n_a(\mu_a,T)&=\frac{1}{\mathcal{A}}\sum_{\bk\sigma}f^a(E^a_{\bk\sigma})\\
  &+\frac{1}{\mathcal{A}}\sum_{b\neq a}\sum_{\sigma\sigma'}\sum_{\nu\bQ}n^B_{ab}(E^{\sigma\sigma'}_{\nu\bQ})\\
  &=n_{\textrm{free}}^{\textrm{GW},a}+n_X
     \,.
    \label{eq:density_final}
\end{split}
\end{equation}
The total carrier density separates into contributions from quasi-free carriers and from carriers bound as excitons according to the two poles in the spectral 
function $A^a(\omega)$. For a specific material, the ionization equilibrium has to be computed numerically. 
The electron and hole chemical potentials are determined by adapting the Fermi functions $f^a(E^a_{\bk\sigma})$ of electrons and holes to a given density of 
quasi-free carriers at the quasi-particle energies $E^a_{\bk\sigma}$. As the chemical potentials also enter the bound-carrier density via the Bose function $n^B_{ab}$, 
Eq.~(\ref{eq:density_final}) represents an implicit equation for the fraction of 
quasi-free carriers $\alpha_a=n_{\textrm{free}}^{a}/n_a$, that has to be solved self-consistently with the quasi-particle 
energies in GW approximation, see Eq.~(\ref{eq:GW_energy}), and the bound-state energies $E^{\sigma\sigma'}_{\nu\bQ}$.
To simplify the procedure, we exploit the fact that shifts of excitonic resonances are naturally much smaller than band-gap shifts, which is due to compensation effects between 
gap shrinkage and binding-energy reduction. \cite{bornath_two-particle_1999,steinhoff_influence_2014} Hence we assume that the exciton spectrum depends only weakly on the 
excitation density so that we can limit ourselves to the BSE (\ref{eq:bse}) in the limit of zero excitation density.

Consistent with the imaginary part of the self-energy (\ref{eq:im_self_en_screened_ladder}), the quasi-particle energies $E^a_{\bk\sigma}$ contain GW- and T-matrix contributions:
% As described in \cite{kremp_quantum_1993, semkat_ionization_2009},
% the Fermi function is expanded around the quasi-particle energy $e^a_{\bk\sigma}$ in GW approximation, which follows from the Dyson equation for the retarded 
% single-particle Green's function using the generalized Kadanoff-Baym ansatz \cite{kremp_quantum_2005}:
%
\begin{equation}
\begin{split} 
  E^a_{\bk\sigma}&=\varepsilon_{\bk\sigma}^a+\textrm{Re}\,\Sigma_{\bk\sigma}^{\textrm{GW},\textrm{ret},a}(E^a_{\bk\sigma})+\textrm{Re}\,\Sigma_{\bk\sigma}^{\textrm{T},\textrm{ret},a}(E^a_{\bk\sigma})
     \,.
    \label{eq:GW_energy_full}
\end{split}
\end{equation}
The GW self-energy is separated into the Fock term and the so-called Montroll-Ward term containing all contributions beyond bare exchange interaction. The T-matrix contribution is explicitely given in
Ref.~\citenum{stolz_correlated_1979} and leads to a blue shift of single-particle energies that is in the nondegenerate case ($f^a(E^a_{\bk\sigma})\ll 1$) caused by the bound-carrier population. 
At the same time, the Fock self-energy contains exchange interaction with both quasi-free and bound carriers via the extended spectral functions that leads to a lowering 
of single-particle energies. 
% Taking a closer look at the Fock self-energy contributing to the quasi-particle energy (\ref{eq:GW_energy}), 
% we find that it does not only contain exchange interaction with quasi-free carriers but also with carriers bound as excitons
% via the spectral function (\ref{eq:spectral_fct_extended}). 
This can be seen by using the T-matrix self-energy in Eq.~(\ref{eq:im_self_en_screened_ladder}) to obtain an excitonic contribution to the spectral function (\ref{eq:spectral_fct_extended}) given by
\begin{equation}
\begin{split} 
  &\Gamma_{\bk\sigma}^a(\omega)= \frac{1}{\mathcal{A}}\sum_{b\neq a} \sum_{\nu\sigma'\bQ}\sum_{\bk'}\\
  &(|\psi^{\sigma\sigma'}_{\nu\bQ}(\bk) |^2\delta_{a,h}\delta_{\bk',\bk+\bQ} + |\psi^{\sigma\sigma'}_{\nu\bQ}(\bk-\bQ) |^2\delta_{a,e}\delta_{\bk',\bk-\bQ}) \\
   &\delta(\hbar\omega+E^b_{\bk'\sigma'}-E^{\sigma\sigma'}_{\nu\bQ})[n^B_{ab}(E^{\sigma\sigma'}_{\nu\bQ})+f^{b}(E^b_{\bk'\sigma'})]
  ~.
    \label{eq:spectral_fct_extended_T_matrix}
\end{split}
\end{equation}
It yields a sharp resonance for each bound state weighted by its Bose population function and the exciton wave functions at the corresponding position in k-space. 
Note that the spectral positions of the resonances are not given by the bound-state energies $E^{\sigma\sigma'}_{\nu\bQ}$, which are two-particle quantities, 
but by an effective binding energy of the carrier in state $\ket{\bk\sigma a}$, as $\Gamma$ represents a single-particle spectral function. 
The Fock self-energy \cite{kremp_quantum_2005} can then be expressed in terms 
of the spectral function using the Kubo-Martin-Schwinger relation for the propagators $G^{<}(\omega)$ in thermal equilibrium:
\begin{equation}
\begin{split} 
    \Sigma_{\bk\sigma}^{\textrm{F},a}=&i\hbar\frac{1}{\mathcal{A}}\sum_{\bk'}V^{aa}_{\bk\bk'\bk\bk'}G^{<,a}_{\bk'\sigma}\\
    =-&i\hbar\frac{1}{\mathcal{A}}\sum_{\bk'}V^{aa}_{\bk\bk'\bk\bk'}\int\frac{d\omega}{2\pi}f^a(\omega)A_{\bk'\sigma}^a(\omega)  \\
    =-&\frac{1}{\mathcal{A}}\sum_{\bk'}V^{aa}_{\bk\bk'\bk\bk'}\Big(f^a(E^a_{\bk'\sigma})+f^{a,\textrm{bound}}_{\bk'\sigma}\Big)\,~.
%     \\
%     -&\frac{1}{\mathcal{A}^2}\sum_{\bk' b}V^{ab}_{\bk\bk'\bk\bk'}\sum_{c\neq b}\sum_{\nu\sigma'\bQ}\sum_{\bk''}\\
%     (|\psi^{\sigma\sigma'}_{\nu\bQ}(\bk') |^2&\delta_{b,h}\delta_{\bk'',\bk'+\bQ} + |\psi^{\sigma\sigma'}_{\nu\bQ}(\bk'-\bQ) |^2\delta_{b,e}\delta_{\bk'',\bk'-\bQ}) \\
%     & n_{bc}(E^{\sigma\sigma'}_{\nu\bQ})[1-f^{b}_{\bk'\sigma}-f^{c}_{\bk'',\sigma'}] \,~.
    \label{eq:HF_self_energy}
\end{split}
\end{equation}
%
% The evaluation of the spectral function runs along the same lines as for the carrier density (\ref{eq:density_full}). It turns out that similar to exchange interaction 
% with free carriers, bound-carrier exchange leads to k-dependent renormalizations according to the exciton wave functions and populations. Note that the population inversion factor 
% $[1-f^{b}_{\bk'\sigma}-f^{c}_{\bk'',\sigma'}]$ is consistent with the normalization of exciton wave functions in the statically screened ladder approximation \cite{stolz_correlated_1979}. 
The first contribution to the Fock self-energy scales, besides the Coulomb matrix elements, with the free-carrier density while the second contribution scales with the density
of bound carriers. It turns out that similar to exchange interaction with free carriers, bound-carrier exchange leads to k-dependent renormalizations according to the exciton wave functions and populations
that are contained in the population factor $f^{a,\textrm{bound}}_{\bk'\sigma}$. As a conclusion, the real part of the self-energy (\ref{eq:GW_energy_full}) contains quasi-particle renormalizations
due to exciton populations via the T-matrix in two different places that act in opposite directions.
We assume that these renormalizations cancel to a large degree and focus on the free-carrier contributions in accordance with Refs.~\citenum{kremp_quantum_1993,semkat_ionization_2009}.
Then we obtain for the quasi-particle energies:
\begin{equation}
\begin{split} 
  E^a_{\bk\sigma}&\approx\varepsilon_{\bk\sigma}^a+\textrm{Re}\,\Sigma_{\bk\sigma}^{\textrm{GW},\textrm{ret},a}(E^a_{\bk\sigma})\Big|_{\textrm{free}}\\
            &=\varepsilon_{\bk\sigma}^a+\Sigma_{\bk\sigma}^{\textrm{F},a}\Big|_{\textrm{free}}+i\hbar\int_{-\infty}^{\infty}\frac{d\omega'}{2\pi}\\
            \sum_{\bq}&\frac{(1-f^a(E^a_{\bq,\sigma}))V^{\textrm{S},>,aa}_{\bk\bq\bk\bq}(\omega')+ 
            f^a(E^a_{\bq,\sigma})V^{\textrm{S},<,aa}_{\bk\bq\bk\bq}(\omega')}{E^a_{\bk\sigma}-E^a_{\bq,\sigma}-\hbar\omega'}
     \,.
    \label{eq:GW_energy}
\end{split}
\end{equation}
In a similar manner as for the Fock self-energy, the full spectral functions could be used to evaluate the Montroll-Ward self-energy in Eq.~(\ref{eq:GW_energy}). 
Due to the spectral structure of the self-energy, however, renormalizations of the single-particle band structure caused by bound carriers involve a denominator of the order of the exciton 
binding energy, which is very off-resonant. Therefore the Montroll-Ward term is evaluated using spectral functions for quasi-free carriers.

\subsection{Screening due to Excited Carriers.}

In the spirit of the extended quasi-particle approximation, screening of the Coulomb interaction due to both free carriers 
and bound excitons is taken into account. The free-carrier screening is treated in RPA with a Lindhard dielectric function \cite{steinhoff_influence_2014}, 
while the excitonic polarizibilities are calculated as described in \cite{ropke_influence_1979,ropke_greens_1981}:
\begin{equation}
\begin{split} 
  &\varepsilon^{\textrm{ret}}_{\bq}(\omega)=\varepsilon^{\textrm{ret},\textrm{RPA}}_{\bq}(\omega)\\
  &-V_{\bq}\frac{1}{\mathcal{A}}\sum_{\nu\nu'\sigma\sigma'\bQ}\left|M^{\sigma\sigma'}_{\nu\nu'\bQ}(\bq)\right|^2\frac{n^B_{ab}(E^{\sigma\sigma'}_{\nu\bQ})-n^B_{ab}(E^{\sigma\sigma'}_{\nu'\bQ-\bq})}
  {E^{\sigma\sigma'}_{\nu\bQ}-E^{\sigma\sigma'}_{\nu'\bQ-\bq}-\hbar\omega-i\gamma}
    \label{eq:screening}
\end{split}
\end{equation}
with matrix elements
\begin{equation}
\begin{split}
&M^{\sigma\sigma'}_{\nu\nu'\bQ}(\bq)=\\
&\frac{1}{\mathcal{A}}\sum_{\bp}\psi^{\sigma\sigma'}_{\nu\bQ}(\bp)\Big[(\psi^{\sigma\sigma'}_{\nu'\bQ-\bq}(\bp+\bq))^*-(\psi^{\sigma\sigma'}_{\nu'\bQ-\bq}(\bp))^*\Big] 
    \label{eq:screening_ME}
\end{split}
\end{equation}
and exciton wave functions $\psi^{\sigma\sigma'}_{\nu\bQ}(\bp)$ as defined above. The momentum and frequency dependence of screening is characterized
by the plasmon spectral function
\begin{equation}
\begin{split} 
\hat{V}^{\textrm{S},ab}_{\bk\bk'\bk\bk'}(\omega)&=V^{\textrm{S},>,ab}_{\bk\bk'\bk\bk'}(\omega)-V^{\textrm{S},<,ab}_{\bk\bk'\bk\bk'}(\omega) \\
&=V^{ab}_{\bk\bk'\bk\bk'}2i\textrm{Im}\,\varepsilon^{-1,\textrm{ret}}_{\bk-\bk'}(\omega)\,.
    \label{eq:W_spec}
\end{split}
\end{equation}

\subsection{Ab initio Single- and Two-Particle Properties.}

% With the theory of ionization equilibrium and bound states outlined in the previous sections at hand we are able to study the balance between
% electron-hole plasma and excitons in various TMDC materials under different experimental conditions. 
%
\begin{figure*}[h!t]
\begin{center}
\includegraphics[width=\textwidth]{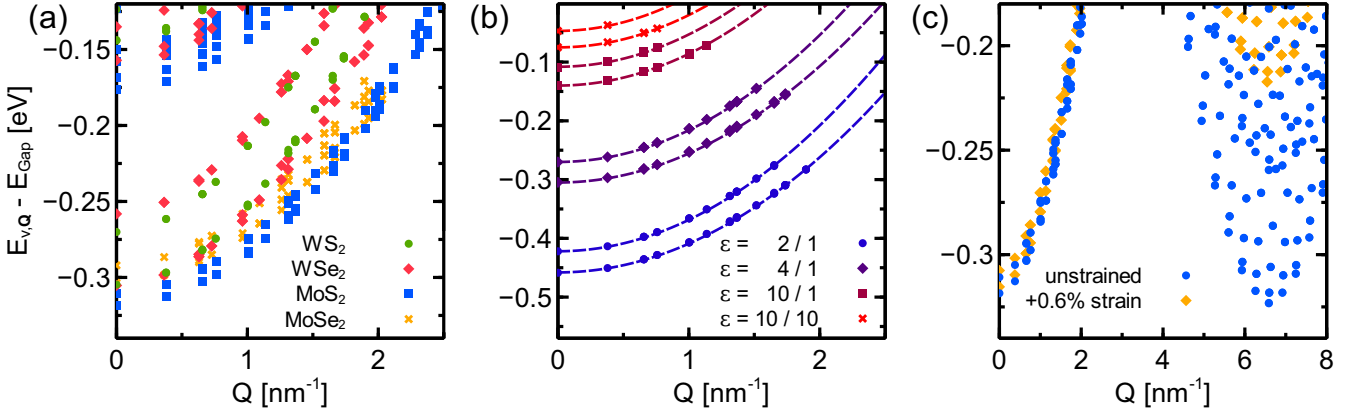}
\caption{\textbf{Bound-state spectra at zero excitation density.} The spectra are measured relative to the quasi-particle band gap and shown 
over the modulus of total exciton momentum $Q$.
\textbf{(a)} Comparison of different TMDC materials on SiO$_2$ substrate. \textbf{(b)} Comparison of WS$_2$ in different dielectric environments with dielectric constant $\varepsilon$ of 
bottom/top environment. Only 1s-excitons are shown. The lines serve as guide to the eye.
\textbf{(c)} Comparison of MoS$_2$ with different lattice constants $a$ on SiO$_2$ substrate. The unstrained layer corresponds to $a=3.18$~\AA\, 
while tensile biaxial strain is simulated by $a=3.20$~\AA\,.}
\label{fig:bound_states}
\end{center}
\end{figure*}
For a material-realistic description of the strong Coulomb interaction in atomically thin systems, we use band structures $\varepsilon^{a}$ and Coulomb matrix elements $V^{ab}$
from material-realistic G$_0$W$_0$ calculations for freestanding slabs of various monolayer TMDCs as a basis, see the Supporting Information and Ref.~\citenum{steinhoff_influence_2014}. 
The matrix elements already take into account dielectric screening due to charge carriers in the ground state of the system. 
% In most experiments the TMDC is placed on top of a substrate. 
Furthermore, we account for additional screening provided by a possible dielectric environment, like a substrate, as described in the Supporting Information. 
The main effect of the additional screening is a reduction of exciton binding energies, which are key quantities for the exciton-plasma balance. 
%As we have in mind experimental or device-related conditions where the TMDC
%is placed on top of a substrate or sandwiched between layers of different materials, we have to think about the modifications of the 2d-layer properties due to the environment.
%We assume that the surrounding materials interact only weakly with the monolayer such that their influence on the band structure is negligible. They mainly affect the internal 
%Coulomb interaction by additional dielectric screening leading to a reduction of exciton binding energies, which are key quantities for the exciton-plasma balance. 
%This screening effect is modelled as described in the Supporting Information. 
The corresponding reduction of quasi-particle band gaps on the other hand plays 
no role here, as all bound-state energies and excited-carrier chemical potentials are measured relative to the gap. By convention, we choose the gap between
the spin-up conduction and valence bands at $K$ as reference which belong to the A-exciton resonance. 
% The valence- and conduction-band splitting caused 
% by spin-orbit interaction is described along the lines of \cite{liu_three-band_2013,steinhoff_influence_2014} including first- and second-order effects. 

% Besides the single-particle band structure and Coulomb matrix elements discussed above 
The calculation of ionization equilibrium requires knowledge of two-particle energies, 
which are obtained by solving the Bethe-Salpeter equation~(\ref{eq:bse}). % and presented in Figs.~\ref{fig:spectralf}~and~\ref{fig:bound_states}.
Fig.~\ref{fig:spectralf}~(b) shows the full exciton spectrum for monolayer WS$_2$ on a SiO$_2$ substrate, for which we assume a dielectric constant $\varepsilon=3.9$. Besides the $K$-excitons 
at $Q=0$ the spectra exhibit a rich structure of various inter-valley excitons with large total momenta. Even for direct-band-gap materials, inter-valley excitons between $K$- and $\Sigma$-valleys
at $Q\approx6$ nm$^{-1}$ may be lower than intra-valley $K$-excitons due to a larger exciton binding energy. 
%The most extreme case is MoSe$_2$, which is already indirect in the single-particle picture 
%according to our calculations. 
As can be seen in Fig.~\ref{fig:bound_states}~(a) the binding energies of the lowest 
$Q=0$-excitons (``1s'') are comparable for all considered materials, the main difference being the splitting between excitons involving different conduction bands, 
which is larger for tungsten-based materials. For MoS$_2$ bound states belonging to the B exciton are visible at $-175$ meV corresponding to the valence-band splitting of
about $140$ meV. Higher exciton states (``2s'') appear as well. Fig.~\ref{fig:bound_states}~(b) shows that dielectric screening from the environment has a strong impact on the exciton spectrum as it
can reduce exciton binding energies significantly \cite{latini_excitons_2015}. This has a major influence on the ionization equilibrium, as we discuss in the main text. Another tunable parameter in experiments
is the application of strain to the TMDC monolayer, which leads to changes in the band structure \cite{conley_bandgap_2013}. This is reflected by MoS$_2$ 
changing from an indirect to a direct semiconductor in the exciton picture under tensile strain, see Fig.~\ref{fig:bound_states}~(c).

\subsection{Numerical Details.}

We calculate the ionization equilibrium from the fraction of quasi-free carriers as root of the implicit equation~(\ref{eq:density_final}).
The two highest valence and two lowest conduction bands are considered to cover all excitons that are relevant in a quasi-equilibrium
situation. Likewise, we limit the Brillouin zone to circles with radius $2.7$ nm$^{-1}$ around the $K$,$K'$,$\Sigma$,$\Sigma'$ and $\Gamma$ points using a Monkhorst-Pack mesh
with $30$ mesh points along $\Gamma$-$M$, which yields reasonable convergence of all results. The frequency integrals involved in the Montroll-Ward self-energy (\ref{eq:GW_energy})
are extended from $-600$ to $600$ meV exploiting the relation $V^{\textrm{S},<,ab}_{\bk\bk'\bk\bk'}(-\omega)=V^{\textrm{S},>,ab}_{\bk\bk'\bk\bk'}(\omega)$.  
For simplicity, we use a dielectric function (\ref{eq:screening}) which is isotropic in momentum by evaluating its dependence on $|\bq|$ along the contour $\Gamma$-$K$ 
and using Coulomb matrix elements $V_{|\bq|}$ that are averaged over Wannier orbitals. Both the Lindhard and the excitonic dielectric function (\ref{eq:screening})
are evaluated using ground-state energies and extrapolated to the limit of vanishing phenomenological quasi-particle broadening $\gamma$.
For a sum over bound states, only those states that are below the local (in $\bk$-space) renormalized band structure measured at the maximum of the corresponding 
exciton wave functions are taken into account. 
% Thus we mimic the vanishing of bound states into the band-edge continuum, which is otherwise not included in the chemical picture of ionization
% equilibrium.

%\begin{acknowledgement}

\textbf{Acknowledgement}
We acknowledge financial support from the Deutsche Forschungsgemeinschaft (JA 14-1 and RTG 2247 "Quantum Mechanical Materials Modelling") and the European Graphene Flagship 
as well as resources for computational time at the HLRN (Hannover/Berlin). We thank Michael Lorke, Christopher Gies and Dirk Semkat for fruitful discussions.

%\end{acknowledgement}

\textbf{Supporting Information available}, containing a description of the ab initio procedures to obtain band structures and screened Coulomb matrix elements.

\section{Supporting Information.}

\renewcommand{\thefigure}{S\arabic{figure}}
\renewcommand{\theequation}{S\arabic{equation}}

\subsection*{Ab-initio based parametrization of the TMDCs \label{app:modeling}}

\begin{figure}
\centering
\includegraphics[width=0.5\textwidth]{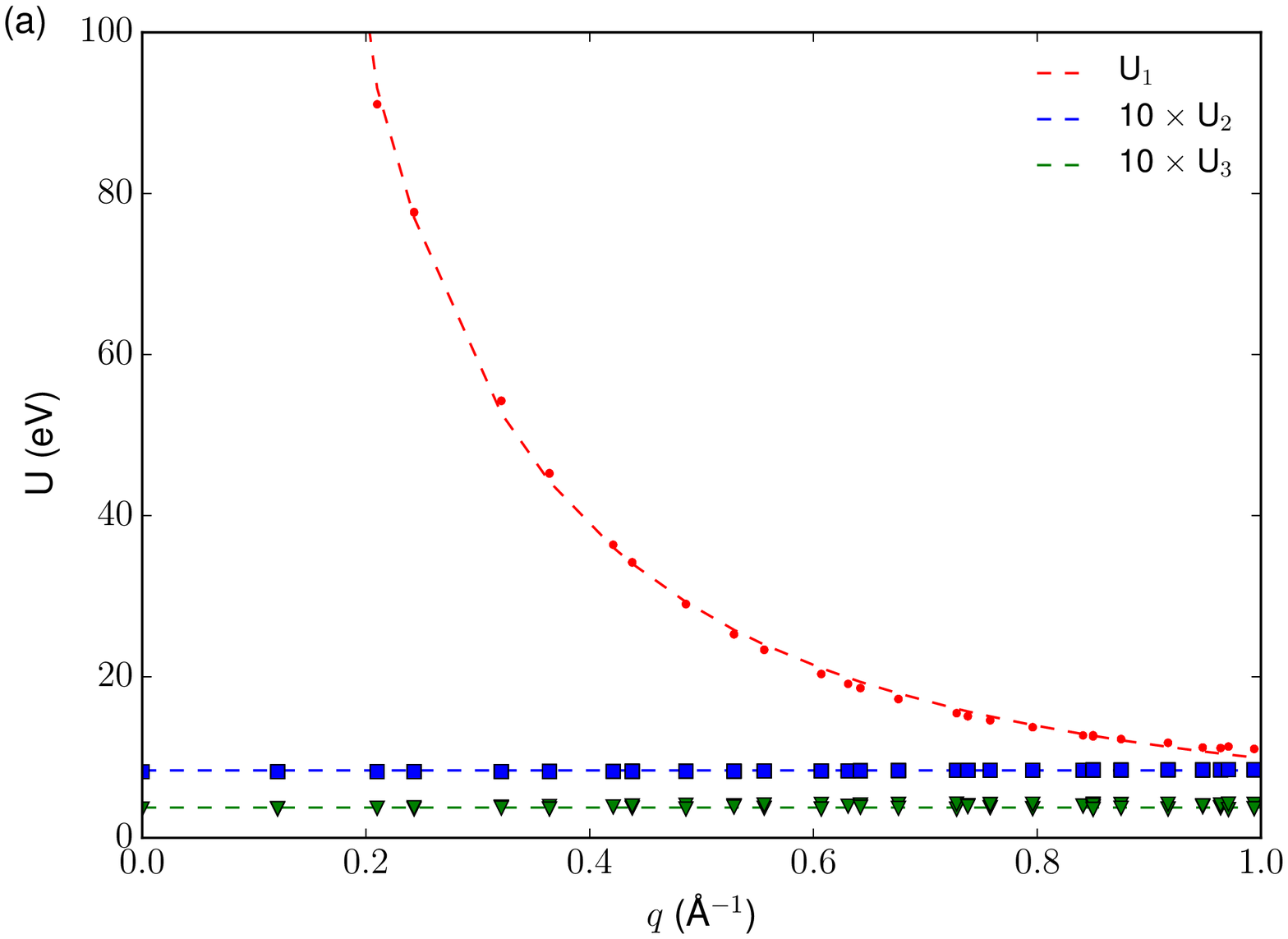}
\includegraphics[width=0.5\textwidth]{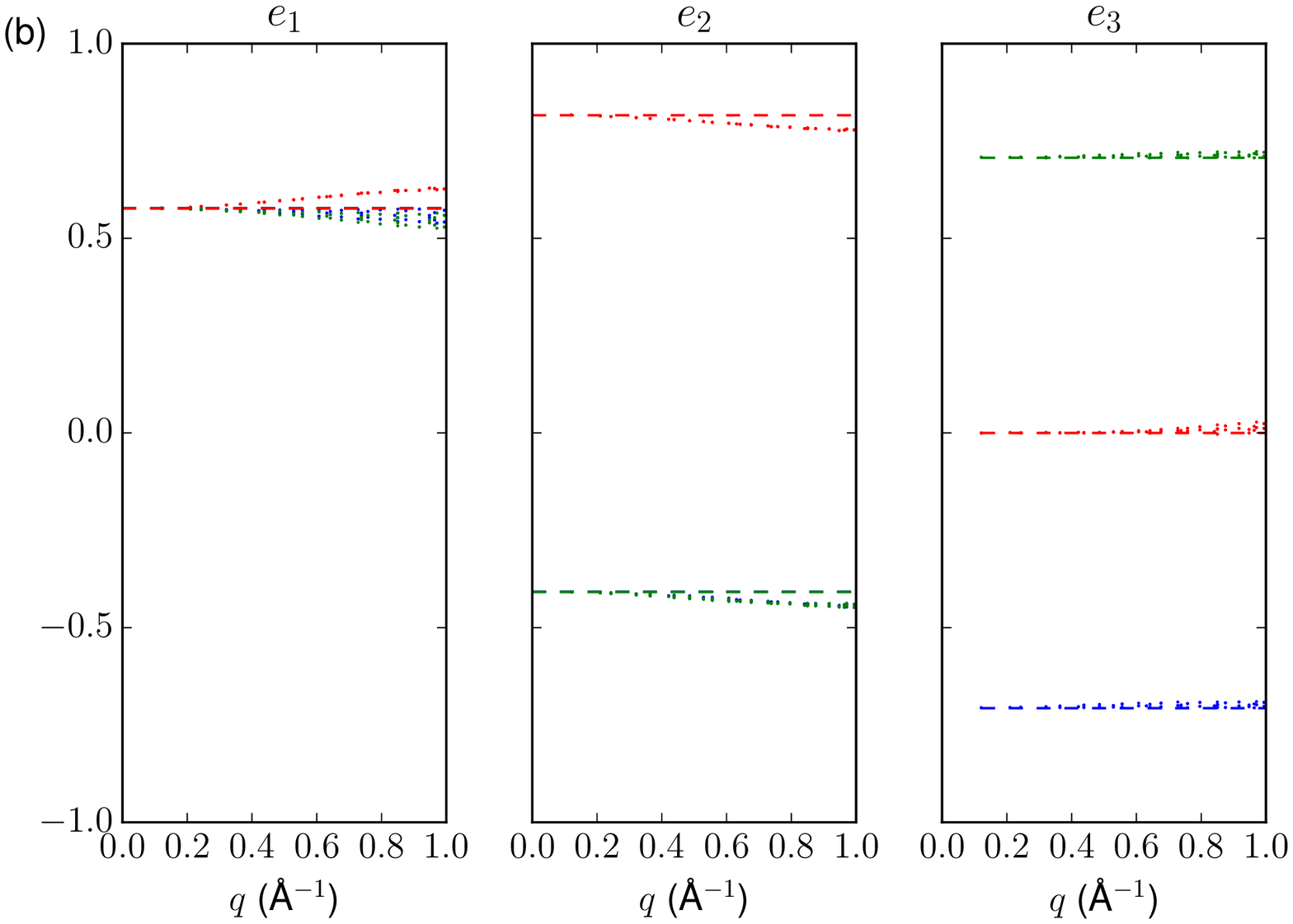}
\includegraphics[width=0.5\textwidth]{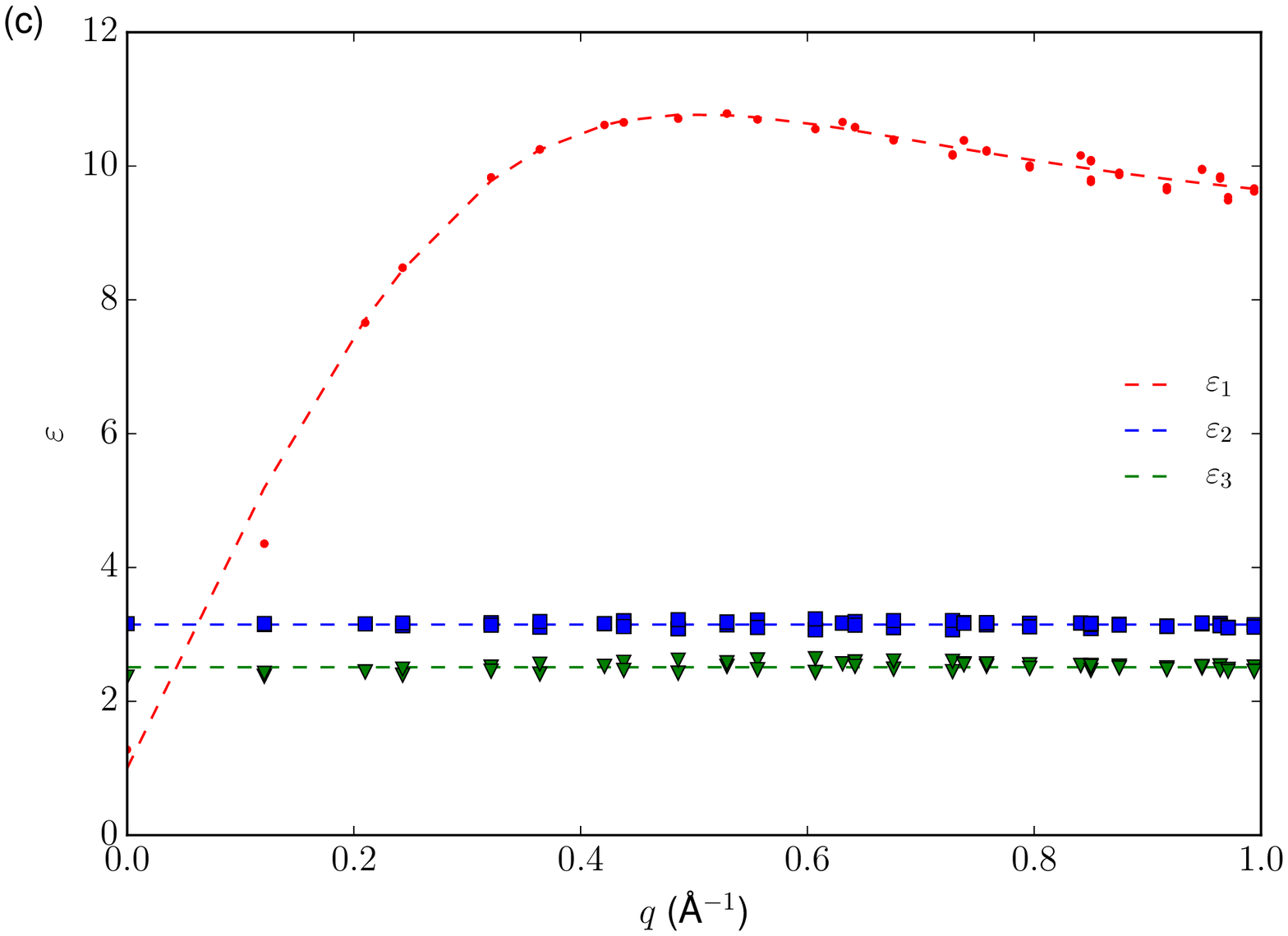}
\caption{Examplary data for the fit and ab initio calculation of the Coulomb interaction in MoSe$_2$. \textbf{(a)} Bare Coulomb matrix elements in its eigenbasis. Red dots, blue squares and green triangles correspond to the leading, second and third eigenvalue of $U(q)$ as obtained from ab initio calculations. Dashed lines show the corresponding fits using Eq. (\ref{eq:v1}) and Tab. \ref{tab:Fits}. \textbf{(b)} Eigenvectors of the bare Coulomb matrix (from left to right corresponding to the leading, second and third eigenvalue). The corresponding vector elements of the $d_{z^2}$ (red), $d_{xy}$ (green) and $d_{x^2-y^2}$ (blue) orbitals are shown. Dashed lines indicate constant analytic values for $U(q\rightarrow 0)$, see Eq. (\ref{eq:vEV}). \textbf{(c)} Matrix elements of the diagonal dielectric function. Markers indicate ab initio results and dashed lines show the fits using Eq. \ref{eq:BackgroundScreening} and Tab. \ref{tab:Fits}.}
\label{fig:CompSpexFit}
\end{figure}

To get an \emph{analytic description} of the Coulomb interaction in semiconducting TMDC monolayers, we apply the approach as presented in \cite{steinhoff_influence_2014,PhysRevB.88.085433,schonhoff_interplay_2016} for MoS$_2$ also to MoSe$_2$, WS$_2$, and WSe$_2$. We start with ab initio calculations for density-density like bare $U_{\alpha\beta} (q)$ and screened $V_{\alpha\beta}(q)$ Coulomb interaction matrix elements in the Wannier basis (with $\alpha, \beta \in [d_{z^2}, d_{xy}, d_{x^2-y^2}]$ ) for the freestanding, undoped TMDC slabs using the FLEUR and SPEX codes \cite{FLEUR,friedrich_efficient_2009,friedrich_efficient_2010} on discrete $18 \times 18 \times 1$ q grids. To interpolate the resulting $3 \times 3$ matrices analytically we make use of the (sorted) eigenbasis of the bare Coulomb interaction $\mathbf{U}$ by diagonalizing it
\begin{align}
\mathbf{U}_\mathrm{diag}(q) = 
  \left(
    \begin{array}{ccc}
      U_1(q) & 0 & 0\\
      0 & U_2 & 0 \\
      0 & 0 & U_3
    \end{array} 
  \right),
\label{eq:BareCoulombInteractionMatrixEigenbasis}
\end{align}
where the diagonal matrix elements are given by
\begin{equation}
U_i = \braket{ e_i | \mathbf{U} | e_i }
\end{equation}
using the eigenvectors of $\mathbf{U}$ in their long-wavelength limits
\begin{equation}
e_1 = \frac{1}{\sqrt{3}} \begin{pmatrix}  1 \\  1 \\  1 \end{pmatrix}, 
e_2 = \frac{1}{\sqrt{6}} \begin{pmatrix} +2 \\ -1 \\ -1\end{pmatrix},
e_3 = \frac{1}{\sqrt{2}} \begin{pmatrix}  0 \\ +1 \\ -1 \end{pmatrix}.
\label{eq:vEV}
\end{equation}
While the leading eigenvalue $U_1(q)$ is a function of $q$ the other two eigenvalues can be readily approximated as constants (see Fig. \ref{fig:CompSpexFit}). For the analytic description of the former we use
\begin{equation}
U_1(q) = \frac{3 e^2}{2 \varepsilon_0 A} \frac{1}{q(1 + \gamma q + \delta q^2)},
\label{eq:v1}
\end{equation}
where $A=\frac{\sqrt{3}}{2}a^2$ is the area of the hexagonal unit cell, $a$ is the lattice constant, and $\varepsilon_0$ is the vacuum permittivity. 

The matrix elements of the screened interaction $\mathbf{V}(q)$ in the eigenbasis of the bare interaction $\mathbf{U}(q)$ are then obtained via 
\begin{equation}
V_i(q) = \varepsilon_i^{-1}(q) \ U_i(q)
\label{eq:CoulombInteractionPartiallyScreened}
\end{equation}
where $\varepsilon_i(q)$ accounts for both the material-specific internal polarizability and the screening by the environment. Its diagonal representation is given by
\begin{equation}
\mathbf{\varepsilon}_\mathrm{diag}(q) = 
  \left(
    \begin{array}{ccc}
      \varepsilon_1(q) & 0 & 0\\
      0 & \varepsilon_2 & 0 \\
      0 & 0 & \varepsilon_3 
    \end{array} 
  \right).
\end{equation}
Once again, the leading eigenvalue $\varepsilon_1(q)$ is a function of $q$ while the other elements are described sufficiently well as constants. Here, the former can be expressed by  
\begin{equation}
\varepsilon_1(q) = 
  \varepsilon_\infty (q)
    \frac{ 1 - \beta_1 \beta_2 e^{-2q\,h}}
       { 1 + (\beta_1 + \beta_2)  e^{-q\,h} + \beta_1 \beta_2 e^{-2q\,h}}
\label{eq:BackgroundScreening}
\end{equation}
which describes the macroscopic dielectric function of a two-dimensional semiconductor. The parameters $\beta_i$ include the screening effects of substrates (see Ref. \cite{rosner_wannier_2015}) via
\begin{equation}
\beta_i = \frac{\varepsilon_\infty (q) - \varepsilon_{\mathrm{sub},i}}{\varepsilon_\infty (q) + \varepsilon_{\mathrm{sub},i}}
\end{equation}
where the dielectric constants of the substrate ($i=1$) and the superstrate ($i=2$) are introduced. In order to describe the original ab initio data as close as possible we fit $\varepsilon_\infty (q)$ using
\begin{equation}
\varepsilon_\infty (q)= \frac{a + q^2}{\frac{a \sin(q c)}{q b c} + q^2 } + e
\label{eq:EpsInf}
\end{equation}
and $\varepsilon_{\mathrm{sub},1}=\varepsilon_{\mathrm{sub},2}=1$ as the ab intio calculations were performed for freestanding layers. 
    
As soon as all fitting parameters are obtained (see Tab. \ref{tab:Fits}) the screening of a dielectric environment can be included by choosing $\varepsilon_{\mathrm{sub},1}$ or $\varepsilon_{\mathrm{sub},2}$ correspondingly. Thus we have a closed analytic description of the screened Coulomb interaction $\mathbf{V}(q)$ in the eigenbasis of the bare interaction $\mathbf{U}(q)$ at arbitrary momenta $q$ in the first Brillouin zone. In order to transform it to the original orbital basis we make use of the eigensystem given in Eq. (\ref{eq:vEV}). In fact, this model is appropriate for every two-dimensional semiconductor.

Besides the analytical description of the screened Coulomb matrix elements we make use of a tight-binding Hamiltonian to describe the electronic band structure (as obtained from $G_0 W_0$ calculations) of the TMDC slab. To this end, we utilize the same Wannier basis as described before (see Ref. \cite{groenewald_valley_2016}) and derive a minimal three-band model describing the highest valence and two lowest conduction bands using the Wannier90 package \cite{mostofi_updated_2014}. Thereby we solely disentangle our target bands from the rest without performing a maximal localization in order to preserve the original transition metal d-orbital characters. The latter is crucial for the subsequent addition of first and  second order Rashba spin-orbit coupling following Ref. \cite{PhysRevB.88.085433}, which takes into account the large spin-orbit splitting in the conduction- and the valence-band K valleys.

Throughout the whole model-building process we assume that the substrate mainly affects the internal Coulomb interaction and neglect its influence on the band structure.
We have used relaxed lattice constants as given in Tab. \ref{tab:Fits}. The values of the bare and screened Coulomb interaction were extrapolated from vacuum heights of $16\,$\AA\ to $32\,$\AA.

\begin{table}[bt]
  \centering
  \caption{Parameters describing the Coulomb interaction and the static screening in the semiconducting TMDCs $MX_2$.}
  \label{tab:Fits}
  \begin{tabular}{|l|cccc|}
    \hline
    & MoS$_2$ & MoSe$_2$ & WS$_2$ & WSe$_2$ \\
    \hline
    lattice constant $a$(\AA ) & 3.180 & 3.320 & 3.191 & 3.325\\
    \hline
    \hline
    bare Interaction $U$ & & & & \\
    \hline
    $\gamma$ (\AA) & 1.932 & 2.232 & 2.130 & 2.297\\
    \hline
    $\delta$ (\AA$^2$) & 0.395 & -0.356 & 0.720 & 0.174\\
    \hline
    A (\AA$^2$) & 8.758 & 9.546 & 8.818 & 9.574 \\
    \hline
    $U_2$ (eV) & 0.810 & 0.837 &  0.712 & 0.715\\
    \hline
    $U_3$ (eV) & 0.367 & 0.376 & 0.354 & 0.360\\
    \hline
    \hline
    screening $\varepsilon$  & & & & \\
    \hline
    $a$ (1/\AA$^2$) & 2.383 & 2.856 & 3.947 & 2.430\\
    \hline
    $b$  & 17.836 & 11.635 & 29.931 & 20.764\\
    \hline
    $c$ (\AA) & 5.107 & 1.979 & 5.440 & 5.761\\
    \hline
    $h$ (\AA) & 2.740 & 4.298 & 1.578 & 2.489\\
    \hline
    $e$  & 5.739 & 6.303 & 4.497 & 5.305\\
    \hline
    $\varepsilon_2$ & 3.077 & 3.148 & 2.979 & 3.028\\
    \hline
    $\varepsilon_3$ & 2.509 & 2.510 & 2.494 & 2.481\\
    \hline
  \end{tabular}
\end{table}

\bibliographystyle{apsrev}

% \bibliography{Ionization_equilibrium}

\end{document}